\documentclass[review]{elsarticle}

\usepackage[top=2.5cm,bottom=2.5cm,left=2.5cm,right=2.5cm]{geometry}
\usepackage{amssymb}
\usepackage{url}
\usepackage{amsfonts,amsmath}
\usepackage{graphicx}
\usepackage{stfloats}

\usepackage[numbers]{natbib}
\usepackage{algorithmic}
\usepackage[ruled,vlined]{algorithm2e}
\usepackage{amsmath, bm}
\usepackage{subcaption}
\usepackage{caption}

\usepackage{multirow}
\usepackage{xfrac}
\usepackage{color}
\usepackage{array}
\usepackage[table]{xcolor}

\usepackage[Symbol]{upgreek}

\usepackage[T1]{fontenc}

\usepackage[utf8]{inputenc}
\usepackage[T1]{fontenc}
\usepackage{amsmath,amssymb,amsfonts}
\usepackage{graphicx}
\usepackage{multirow}
\usepackage{epstopdf} 
\usepackage{scalefnt}
\usepackage{gensymb}

\journal{arXiv.org}

\begin{document}

\begin{frontmatter}

\title{Proposal of a Takagi-Sugeno Fuzzy-PI Controller Hardware}

\author[1]{S\'{e}rgio N. Silva}
\ead{sergionatan@dca.ufrn.br}

\author[1]{Felipe F. Lopes}
\ead{lopesffernandes@gmail.com}

\author[2]{Carlos Valderrama}
\ead{carlosAlberto.VALDERRAMASAKUYAMA@umons.ac.be}

\author[1,3]{Marcelo A. C. Fernandes\corref{cor1}\fnref{ca1}}
\ead{mfernandes@dca.ufrn.br}

\cortext[cor1]{Corresponding author}
\fntext[ca1]{Present address: John A. Paulson School of Engineering and Applied Sciences, Harvard University, Cambridge, MA 02138, USA.}

\address[1]{Laboratory of Machine Learning and Intelligent Instrumentation, Federal University of Rio Grande do Norte, Natal 59078-970, Brazil.}
\address[2]{Department of Electronics \& Microelectronics, Polytechnic Faculty, University of Mons, Mons, 7000, Belgium.}
\address[3]{Department of Computer Engineering and Automation, Federal University of Rio Grande do Norte, Natal, RN, 59078-970, Brazil.}

\begin{abstract}
This work proposes dedicated hardware for an intelligent control system on Field Programmable Gate Array (FPGA). The intelligent system is represented as Takagi-Sugeno Fuzzy-PI controller. The implementation uses a fully parallel strategy associated with a hybrid bit format scheme (fixed-point and other floating-point). Two hardware designs are proposed; the first one uses a single clock cycle processing architecture, and the other uses a pipeline scheme. The bit accuracy was tested by simulation with a non linear control system of robotic manipulator. The area, throughput, and dynamic power consumption of the implemented hardware are used to validate and compare the results of this proposal. The results achieved allow that the proposal hardware can use in several applications with high-throughput, low-power and ultra-low-latency restrictions such as teleportation of robot manipulators, tactile internet, industrial automation in industry 4.0, and others.
\end{abstract}


\end{frontmatter}


\section{Introduction}

Systems based on Fuzzy Logic (FL), have been used in many industrial and commercial applications such as robotics, automation, control and classification problems. Unlike high data volume systems, such as Big Data and Mining of Massive Datasets (MMD) \cite{fuzzyBig1, fuzzyBig2, fuzzyBig3}, one of the great advantages of Fuzzy Logic is its ability to work with incomplete or inaccurate information.

Intelligent systems based on production rules that use Fuzzy Logic in the inference process are called in the literature of Fuzzy Systems (FS) \cite{FuzzyRef}. 
Among the existing inference strategies, the most used, the Mamdani and the Takagi-Sugeno, are differentiated by the final stage of the inference process
\cite{PaperFPGAFuzzySugenoControl1, PaperFPGAFuzzySugenoControl2, PaperFPGAFuzzySugenoControl3, PaperFPGAFuzzySugenoControl4, PaperFPGAFuzzySugenoControl5, PaperFPGAFuzzySugenoControl6, PaperFPGAFuzzySugenoControl7A, PaperFPGAFuzzySugenoControl7B, PaperFPGAFuzzySugenoControl7C, PaperFuzzyControlFPGA1, PaperFuzzyControlFPGA2, PaperFuzzyControlFPGA3, PaperFuzzyControlFPGA5, PaperFuzzyControlFPGA6, PaperFuzzyControlFPGA7, PaperFuzzyControlFPGA8}. 

The interest in the development of dedicated hardware implementing Fuzzy Systems has increased due to the demand for high-throughput, low-power and ultra-low-latency control systems for emerging applications such as the tactile Internet \cite{Tactile1RESRC, tactile}, the Internet of Things (IoT) and Industry 4.0, where the problems associated with processing, power, latency and miniaturization are fundamental. Robotic manipulators used on tactile internet need a high-throughput and ultra-low-latency control system, and this can be achieved with dedicated hardware \cite{Tactile1RESRC}.

The development of dedicated hardware, in addition to speeding up parallel processing, makes it possible to operate with clocks adapted to low-power consumption \cite{torquato2018highperformance, 8626462, silva2019reinforcementlearning, 8678408, BLAIECH2019331, electronics8060631, NORONHA2019138}. The works presented in  \cite{PaperFPGAFuzzy3, PaperFPGAFuzzy4, PaperFPGAFuzzy5, PaperFPGAFuzzy6, PaperFPGAFuzzy7, PaperFPGAFuzzy12, PaperFPGAFuzzySurvey1, PaperFPGAFuzzySurvey2} propose implementations of FS on reconfigurable hardware (Field Programmable Gate Array - FPGA), showing the possibilities associated with the acceleration of fuzzy inference processes having a high degree of parallelization. Other works propose specific implementations of Fuzzy Control Systems (FCS) using the Fuzzy Mamdani Inference Machine (M-FIM) and the Takagi-Sugeno Fuzzy Inference Machine (TS-FIM) \cite{PaperFPGAFuzzySugenoControl1, PaperFPGAFuzzySugenoControl2, PaperFPGAFuzzySugenoControl3, PaperFPGAFuzzySugenoControl4, PaperFPGAFuzzySugenoControl5, PaperFPGAFuzzySugenoControl6, PaperFPGAFuzzySugenoControl7A, PaperFPGAFuzzySugenoControl7B, PaperFPGAFuzzySugenoControl7C, PaperFuzzyControlFPGA1, PaperFuzzyControlFPGA2, PaperFuzzyControlFPGA3, PaperFuzzyControlFPGA5, PaperFuzzyControlFPGA6, PaperFuzzyControlFPGA7, PaperFuzzyControlFPGA8}. The works presented in \cite{PaperFPGAFuzzySugeno1, PaperFPGAFuzzySugeno2, PaperFPGAFuzzySugeno3} propose the Takagi-Sugeno hardware acceleration for other types of application fields. 

This work aims to develop a new hardware proposal for a Fuzzy-PI controller with TS-FIM. Unlike most of the works presented, this project offers a fully parallel scheme associated with a hybrid platform using fixed-point and floating-point representations. Two TS-FIM hardware modules have been proposed, the first (here called TS-FIM module one-shot) takes one sample time to execute the TS-FIM, and the second (here called as TS-FIM module pipeline) uses registers inside the TS-FIM. Two Fuzzy-PI controller hardware have been proposed, one for the TS-FIM one-shot module and another for the TS-FIM module pipeline. The proposed hardware have been implemented, tested and validated on a Xilinx Virtex 6 FPGA. The synthesis results, in terms of size, resources and throughput, are presented according to the number of bits and the type of numerical precision. Already, the physical area on the target FPGA reaches less than $7\%$. The implementation achieved a throughput between $10$ and $18\text{Msps}$ (Mega samples per second), and between $490$ and $882 \, \text{Mflips}$ (Mega fuzzy logic inferences per second). Validation results on a feedback control system are also presented, in which satisfactory performance has been obtained for a small number of representation bits. Comparisons of results with other proposals in the literature in terms of throughput, hardware resources, and dynamic power savings will also be presented. 


\section{Related works}

In \cite{PaperFPGAFuzzy3}, a high-performance FPGA Mamdani fuzzy processor is presented. The processor achieved a throughput of about $5 \, \text{Mflips}$ at a clock frequency about $40 \, \text{MHz}$ and it was designed for $256$ rules and $16$ inputs with $16$ bits. The proposal used a semi-parallel implementation and thus reduced the number of the operations per Hz. The work presented in \cite{PaperFPGAFuzzy3} has about $\frac{5}{40} = 0.125 \, \text{flips/Hz}$ and the work proposed here can achieve about $\frac{256*40}{40} = 256 \, \text{flips/Hz}$ due the fully parallel hardware scheme used. The significant difference between throughput and operation frequency also implies a high power consumption \cite{MCCOOL201239}. The work presented in \cite{PaperFPGAFuzzy4} uses a Mamdani inference machine and the throughput in Mflips is about $48.23\, \text{Mflips}$. The hardware was designed to operate with $8$ bits, four inputs, $9$ rules and one output. Similar to the work presented in \cite{PaperFPGAFuzzy3}, the proposal introduced in \cite{PaperFPGAFuzzy4} adopted a semi-parallel implementation, and this way decreased the throughput and increased power consumption. Other Mamdani implementations following the same strategy are also found in \cite{PaperFPGAFuzzy5, PaperFPGAFuzzy6, PaperFPGAFuzzy7, PaperFPGAFuzzy12}. 

A multivariate Takagi-Sugeno fuzzy controller on FPGA is proposed in \cite{PaperFPGAFuzzySugenoControl1}. The hardware is applied to the temperature and
humidity controller for a chicken incubator and it was projected to two inputs, $6$ rules and three outputs. When compared to other works, the hardware proposed in \cite{PaperFPGAFuzzySugenoControl1} achieved a low throughput about $6 \, \text{Mflips}$. A hardware accelerator architecture for
a Takagi-Sugeno fuzzy controller is proposed in \cite{PaperFPGAFuzzySugenoControl3} and this proposal achieved a throughput about $1.56\, \text{Msps}$ with three inputs, two outputs and $24$ bits.

In \cite{PaperFPGAFuzzySugenoControl7A, PaperFPGAFuzzySugenoControl7B, PaperFPGAFuzzySugenoControl7C} a design methodology for rapid development of fuzzy controllers on FPGAs was developed. For the case with two inputs, $35$ rules and one output (vehicle parking problem), the proposed hardware achieved a maximum clock about $66.251 \, \text{MHz}$ with $10$ bits. However, the TS-FIM takes $10$ clocks to complete the inference step, and this decreases the throughput, and it increases the power consumption.

The implementation presented in \cite{PaperFuzzyControlFPGA1} aims at creating a hardware scheme of fuzzy logic controller on FPGA for the maximum power point tracking in photo-voltaic systems. The implementation takes $6$ clocks cycles over $10 \, \text{MHz}$ and this is equivalent a throughput about $\frac{10 \, \text{MHz}}{6} \approx 1.67 \, \text{Msps}$. In \cite{PaperFuzzyControlFPGA3}, a Mamdani fuzzy logic controller on FPGA was proposed. The hardware carries out a throughput of about $25 \, \text{Mflips}$ with two inputs, $49$ rules. 

The work presented in \cite{PaperFuzzyControlFPGA5} implements a semi-parallel digital fuzzy logic controller on FPGA. The work achieved about $16 \, \text{Msps}$ per clock frequency of $200 \, \text{MHz}$, that is, $0.08 \, \text{Msps/MHz}$. On the other hand, this manuscript uses a fully parallel approach and it achieves $1 \, \text{Msps/MHz}$, in other words, it can execute more operations per clock cycle. In the same direction, the proposals presented in \cite{PaperFuzzyControlFPGA6, PaperFuzzyControlFPGA8} shows a semi-parallel fuzzy control hardware with low-throughput, about $1\, \text{Msps}$. 

Thus, this manuscript proposes a hardware architecture for the Fuzzy-PI control system. Unlike the works presented in the literature, the strategy proposed here uses a fully parallel scheme associated with a hybrid use in the bit format (fixed and floating-point). After several comparisons with other implementations of the literature, the scheme proposed here showed significant gains in processing speed (throughput) and dynamic power savings.

\section{Takagi-Sugeno Fuzzy-PI Controller}\label{sec2}

\begin{figure}[ht]
\begin{center}
  \includegraphics[width=0.98\textwidth]{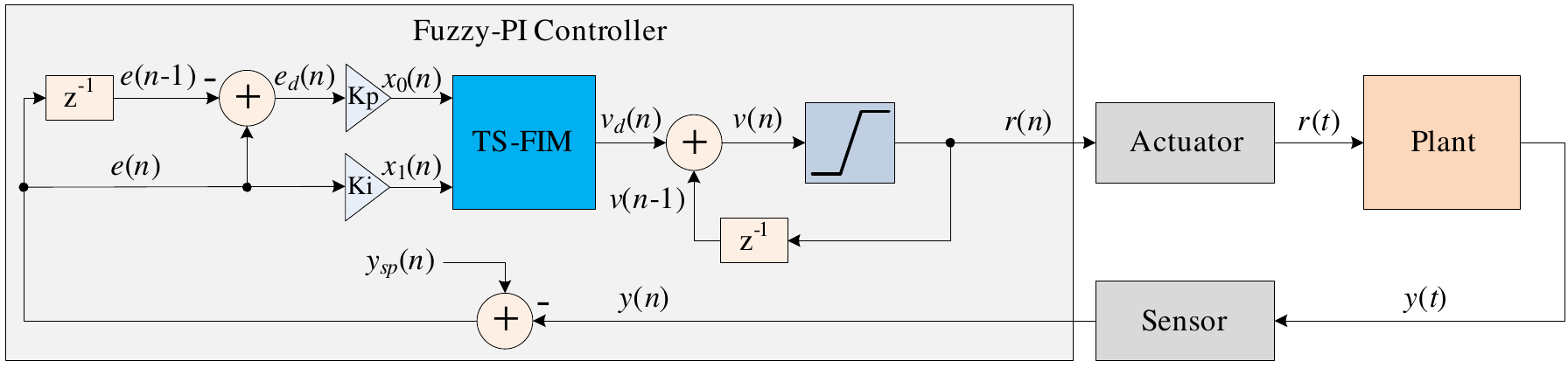}\\
  \caption{Architecture of the Fuzzy-PI feedback control system operating a generic plant.}\label{FuzzyPIController}
\end{center}
\end{figure}

Figure \ref{FuzzyPIController} shows the Fuzzy-PI intelligent control system operating a generic plant \cite{FuzzyRef, Silva2019, FERNANDES2016684}. The plant output variable $y(t)$ is called the controlled variable (or controlled signal), and it can admit several kinds of physical measurements such as level, angular velocity, linear velocity, angle, and others depending on the plant characteristics. The controlled variable, $y(t)$, passes through a sensor that converts the physical measure into a proportional electrical signal that it is discretized at a sampling rate, $t_s$, generating the signal, $y(n)$. 

The plant drives the kind of sensor that will be used. For level control in tanks used in industrial automation, the sensor can be characterized by the pressure sensor. For robotics applications (manipulators or mobile robotics), the sensor can be position sensor (capture angle information) or encoders sensor (capture angular or linear velocity information).

In the $n$-th time, the Fuzzy-PI controller (see Figure \ref{FuzzyPIController}) uses the signal, $y(n)$, and it calculates the error signal, $e(n)$, and difference of error, $e_d(n)$. The signal $e(n)$ is expressed by
\begin{equation}\label{ErrorEq}
e(n)= y_{sp}(n) - y(n),
\end{equation} 
where the $y_{sp}(n)$ is the reference signal also called the set point variable and the signal $e_d(n)$ by
\begin{equation}\label{DiffErrorEq}
e_d(n)= e(n)-e(n-1).
\end{equation} 
After the computation of the signals $e(n)$ and $e_d(n)$, the Fuzzy-PI controller generate the signals $x_1(n)$ and $x_2(n)$, which can be expressed as
\begin{equation}\label{KpEq}
x_0(n)= \text{Kp} \times e_d(n)
\end{equation} 
and
\begin{equation}\label{KiEq}
x_1(n)= \text{Ki} \times e(n).
\end{equation} 
The variables $\text{Kp}$ and $\text{Ki}$ represent the proportional gain and the integration gain, respectively \cite{FuzzyRef, Silva2019, FERNANDES2016684}. Subsequently, the signals $x_0(n)$ and $x_1(n)$ are sent to the fuzzy Takagi-Sugeno inference, called in this article of Takagi-Sugeno - Fuzzy Inference Machine (TS-FIM) (see Figure \ref{FuzzyPIController}). 

The TS-FIM is formed by three stages called fuzzification, operation of the rules (or rules evaluation) and defuzzification (or output function) \cite{FuzzyRef}. In the fuzzification each $i$-th input signal $x_i(n)$ is applied to a set of $F_i$ pertinence functions whose output can be expressed as 
\begin{equation}
f_{i,j}(n)=\mu_{i,j}(x_i(n)) \text{ for } j=0,\dots,F_i,
\end{equation}
where, $\mu_{i,j}(\cdot)$ is the $j$-th membership function of the $i$-th input and $f_{i,j}(n)$ is the output of the fuzzification step associated with the $j$-th membership function and the $i$-th input in the $n$-th time. For two inputs, $x_0(n)$ and $x_1(n)$, the TS-FIM generates a set of $F_0 + F_1$ fuzzy signals ($f_{0, j}$ and $f_{1, j}$) and these signals are processed by a set of $F_0F_1 $ rules in the operation (or evaluation) phase. Each $g$-th rule can be expressed as
\begin{equation}\label{EqOM}
o_{g}=\min(f_{0,l},f_{1,k}) \text{ for } g=0,\dots,\left(F_0F_1\right)-1, 
\end{equation}
where $g = F_{0,l+k}$ for $(l, k)=(0,0), (0,1),\dots,(F_0-1, F_1-1)$. Finally, the output (defuzzification) of TS-FIM, called here $v_d(n) $, can be expressed as
\begin{equation}
v_d(n) = \frac{a(n)}{b(n)}=\frac{\sum_{g=1}^{\left(F_0F_1\right)-1}a_{g}}{\sum_{g=0}^{\left(F_0F_1\right)-1}o_{g}}=\frac{\sum_{g=1}^{\left(F_0F_1\right)-1}o_{g}\times \left(A_gx_0(n) + B_gx_1(n) + C_g \right)}{\sum_{g=0}^{\left(F_0F_1\right)-1}o_{g}},
\label{EqOFM}
\end{equation}
where $A_g$, $B_g$ e $C_g$ are parameters defined during the project \cite{FuzzyRef}.Thus it can be said that every $n$-th instant TS-FIM receives as input $x_0(n)$ and $x_1(n)$ and generates as output $v (n)$, that is,
\begin{equation}\label{EqTSFIM}
v_d(n) = \text{\emph{TSFIM}}\left(x_0(n),x_1(n)\right),
\end{equation}
where $\text{\emph{TSFIM}}\left(\cdot\right)$ is a function that represents TS-FIM.

After the TS-FIM processing, the Fuzzy-PI controller integrates the signal $v_d(n)$ generating the signal $v(n)$ (see Figure \ref{FuzzyPIController}). The signal is the output of the Fuzzy-PI controller, and it can be expressed as
\begin{equation}\label{EqInt}
v(n)= v_d(n) + v(n-1).
\end{equation} 
The signal $v(n)$ is saturated between $v_{\text{min}}$ and $v_{\text{max}}$, generating the signal $r(n)$ that it is expressed as
\begin{equation}\label{EqSat}
r(n)= \left\{ 
\begin{matrix}
v_{\text{max}} & \text{ for } v(n) > v_{\text{max}} \\
v(n) &\\
v_{\text{min}} & \text{ for } v(n) < v_{\text{min}} 
\end{matrix} \right. .
\end{equation}
Finally, the signal $r(n)$ is sent to a actuator, which transforms the discrete signal into a continuous signal, $r(t)$, to be applied to the plant. 

\section{Hardware Proposal}\label{secHP}

\begin{figure}[ht]
	\centering
	\includegraphics[width=1\textwidth]{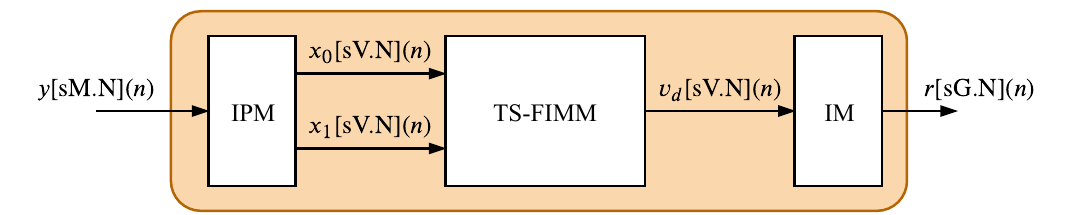}
	\caption{Overview of Fuzzy-PI controller proposed architecture.}
	\label{GeralHardware}
\end{figure}

Figure \ref{GeralHardware} presents the general structure of the proposed hardware in which it consists of three main modules called Input Processing Module (IPM), TS-FIM Module (TS-FIMM) and Integration Module (IM). The hardware was developed for the most part using a fixed-point format for the variables, in which, for any given variable, the notations $[\text{uT.W}]$ and $[\text{sT.W}]$ indicate that the variable is formed by $\text{T}$ bits of which $\text{W}$ are intended for the fractional part and the symbols "$\text{s}$" and "$\text{u}$" indicate that the variable is signed or unsigned, respectively. For the case of signed variables, type $s$, the number of bits destined for the integer part is characterized as $ \text{T} - \text{W} -1$ and for unsigned variables, type $u$, the number of bits is $ \text{T} - \text{W}$ for the integer part.

\subsection{Input Processing Module (IPM)}\label{secIPM}

\begin{figure}[ht]
	\centering
	\includegraphics[width=.95\textwidth]{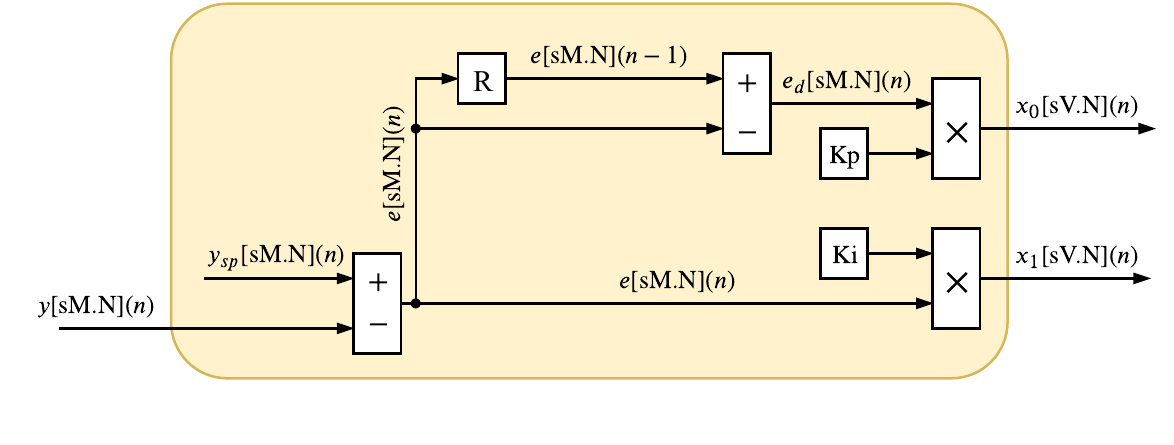}
	\caption{Hardware architecture of IPM.}
	\label{IPMHardware}
\end{figure}

The IPM (shown in Figure \ref{IPMHardware}) is responsible for processing the control signal generated by the plant to the input of the Fuzzy-PI controller. The IPM computes the Equations \ref{ErrorEq}, \ref{DiffErrorEq}, \ref{KpEq} and \ref{KiEq}.  The signals associated with this module were implemented with $M$ bits where, one is reserved for the sign and $N$ for the fractional part where, the value of $M$ can be expressed as
\begin{equation}
M = N + \log_2(\lceil y_{max} \rceil) + 1,
\end{equation}
where $y_{max}$ represents the maximum value, in modulus, of the process variable, $y(n)$. The values of $\text{Kp}$ and $\text{Ki}$ should be designed to try to maintain the output signals of the module, $x_0[\text{V.N}] (n)$ and $x_1[\text{V.N}](n)$, between $-1$ and $1$, respectively. In this way, you can set $\text{V}=\text{N}+1$, aiming at reducing the number of bits associated with the project. It is important to note that the two gain modules, $\text{Kp}$ and $\text{Ki}$, also saturate the signal in $[\text{V.N}](n)$ bits after multiplication.

\subsection{TS-FIM Module (TS-FIMM)}\label{secTSFIMHardware}

The TS-FIMM is composed of three hardware components: Membership Function Module (MFM), Operation Module (OM) and Output Function Module (OFM). The MFM is the first module associated with TS-FIMM and it corresponds to the fuzzification process, the OM component completes the rules evaluation phase and the OFM performs the defuzzification step (see Section \ref{sec2}). This work proposes two designers for TS-FIMM. 

\begin{figure}[ht]
	\centering
	\includegraphics[width=.8\textwidth]{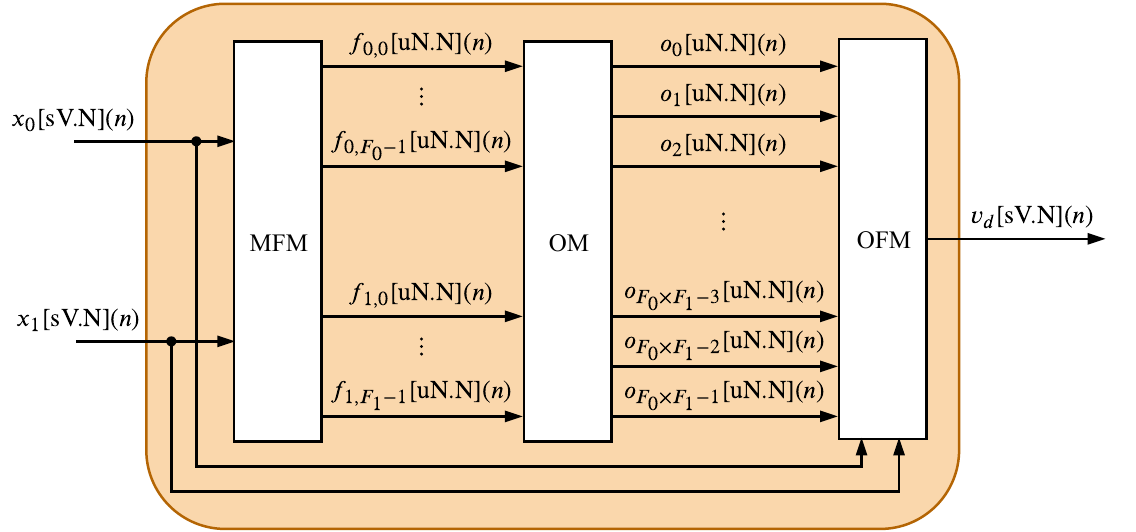}
	\caption{Hardware architecture of TS-FIMM One-Shot (TS-FIMM-OS).}
	\label{TSFIMHardwareP1}
\end{figure}

\begin{figure}[ht]
	\centering
	\includegraphics[width=1\textwidth]{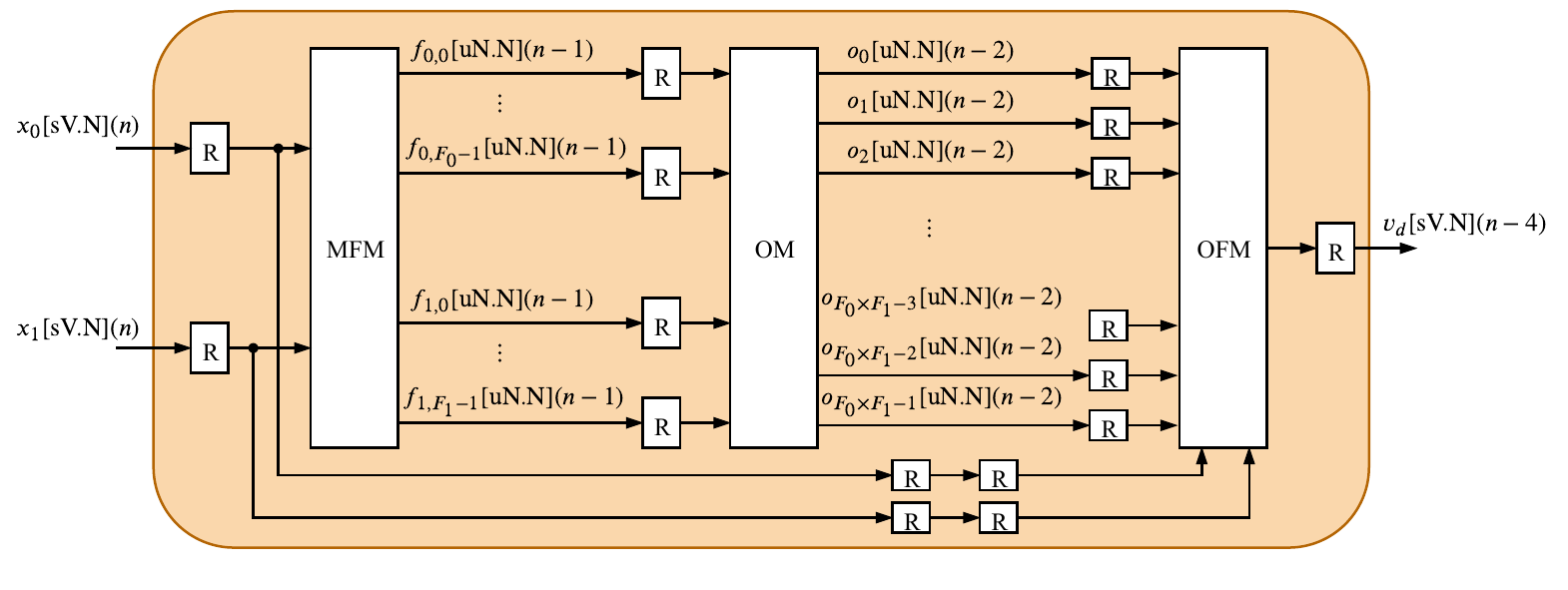}
	\caption{Hardware architecture of TS-FIMM Pipeline (TS-FIMM-P).}
	\label{TSFIMHardwareP2}
\end{figure}

The first one, presented in Figure \ref{TSFIMHardwareP1} and called here as TS-FIMM One-Shot (TS-FIMM-OS), performs all modules MFM, OM, and OFM in one sample time, in other words, it takes one sample time to generate the $n$-th output associated of the $n$-th input. The second, presented in Figure \ref{TSFIMHardwareP2} and called here as TS-FIMM Pipeline (TS-FIMM-P), used registers (blocks called R in the Figure \ref{TSFIMHardwareP1}) among the input, MFM, OM, OFM and output. The TS-FIMM-P takes four sample time to perform all modules MFM, OM, and OFM, in other words, there is a delay of the four samples between the $n$-th output and $n$-th input. 

The TS-FIMM-OS will have a longer sample time than TS-FIMM-P because the critical path is also longer; however, the TS-FIMM-OS does not have a delay. It is important to empathize that the delay inside the feedback control can take a system to instability. The instability degree depends on the system and how long is the delay. The instability will depend on the characteristics of the system and the size of the delay \cite{Niculescu2010}. On the other hand, the pipeline scheme associated with TS-FIMM-P has a short sample time (short critical path), and this permits a high-throughput when it compares to TS-FIMM.  

\subsubsection{Membership Function Module (MFM)}\label{sec3}

\begin{figure}[ht]
    \centering
    \includegraphics[width=0.45\columnwidth]{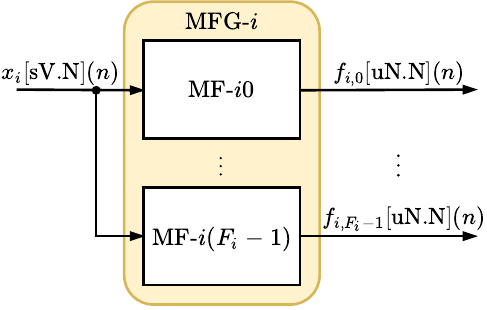}
    \caption{Hardware architecture of module MFG-$i$ associated with the $i$-th input, $x_i[\text{sV.N}](n)$.}
    \label{MFGi}
\end{figure}

In the MFM, each $i$-th input variable is associated with a module that collects $F_i$ membership functions, called here Membership Function Group (MFG). Figure \ref{MFGi} shows the $i$-th MFG, called of the $\text{MFG-}i$, related with the $i$-th input, $x_i[\text{sV.N}](n)$. 

Each $\text{MFG-}i$ collects $F_i$ membership functions (see Figure \ref{MFGi}) called $\text{MF-}ij$ and each module $\text{MF-}ij$ implements the $j$-th membership function associated with the $i$-th input, $\mu_{i, j} (x_i (n))$. In every $n$-th time instant all membership functions, $\sum_i{F_i}$, are executed in parallel and at the output of each $ \text{MF-}ij$ is generated a $N$ bits signal of type $\text{u}$ and without the integer part, called $f_{i, j}[\text{uN.N}] (n)$ (see Figure \ref{MFGi}). The Fuzzy-PI controller proposed here uses $F_0 + F_1$ membership functions.

\begin{figure}[ht]
    \begin{center}
    \includegraphics[width=0.8\textwidth]{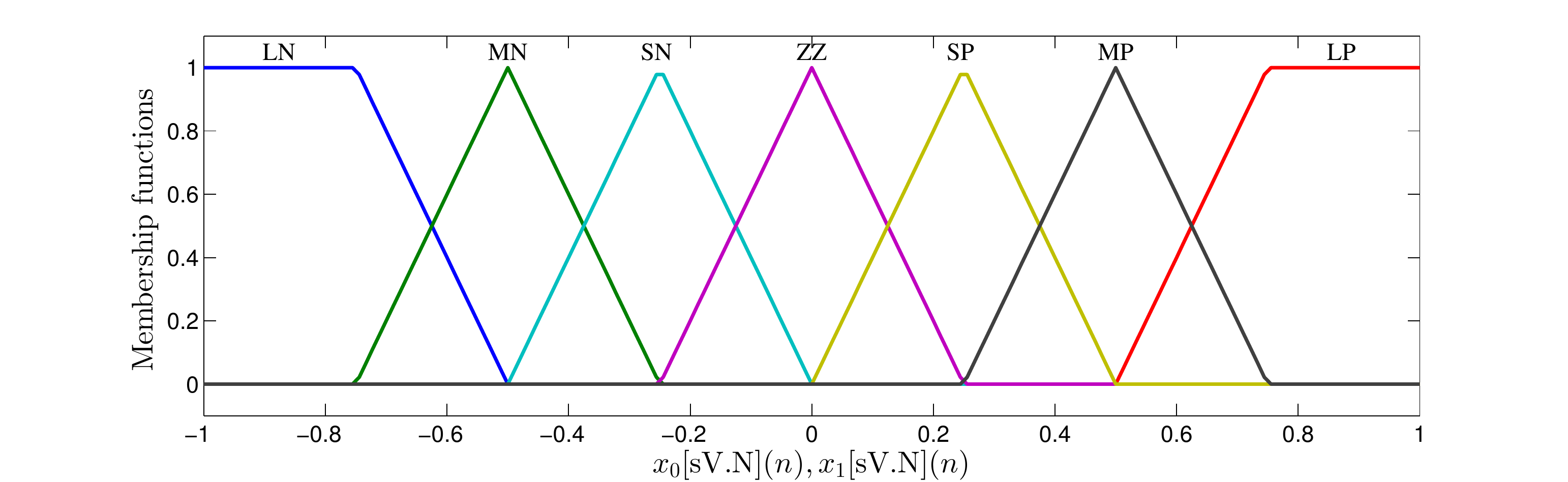}
    \caption{Membership functions from inputs $x_0[\text{sV.N}](n)$ and $x_1[\text{sV.N}](n)$.} \label{fig:MFIn1}
    \end{center}
\end{figure}

Figure \ref{fig:MFIn1} shows the membership functions implemented in the MFM. For both variables, $x_0 [\text{sV.N}] (n) $ and $x_1 [\text{sV.N}] (n)$, seven functions of pertinence were created (trapezoidal type in the extremes and triangular in the remaining). The linguistic terms associated with membership functions are Large Negative (LN), Moderate Negative (MN), Small Negative (SN), Zero (ZZ), Small Positive (SP), Moderate Positive (MP) and Large Positive (LP).

Each $j$-th membership function associated with $i$-th input was implemented directly on hardware based on the following expressions
\begin{eqnarray}
    \mu^{RT}_{i,j}(x_i\text{[sV.N}](n)) =  \left\{ \begin{array}{ll}
    0 & \text{if }  x_i[\text{sV.N}](n) > d_{i,j}[\text{sW.T}] \\
     G^{RT}_{i,j}(n) & \text{if } c_{i,j}[\text{sW.T}] \leq x_i[\text{sV.N}](n) \leq d_{i,j}[\text{sW.T}],\\
    1 & \text{if }  x_i[\text{sV.N}](n) < c_{i,j}[\text{sW.T}] \\
    \end{array}
    \right.
\end{eqnarray}
being $\mu^{RT}_{i,j}(\cdot)$ the trapezoidal function on the right, $c_{i,j}[\text{sW.T}$ and $d_{i,j}[\text{sW.T}]$ are constants ($c_{i,j}[\text{sW.T}$ < $d_{i,j}[\text{sW.T}]$) and
\begin{equation}
    G^{RT}_{ij}(n) = \frac{d_{i,j}[\text{W.T}] - x_i[\text{sV.N}](n)}{d_{i,j}[\text{W.T}] - c_{i,j}[\text{W.T}]},
\end{equation}
where $W$ and $T$ are the number of bits in the integer and fractional part relative to the constants of the $j$-th activation function associated with $i$-th input. For the trapezoidal of the left one has
\begin{eqnarray}
    \mu^{LT}_{i,j}(x_i\text{[sV.N}](n)) =  \left\{ \begin{array}{ll}
    0 & \text{ if }  x_i[\text{sV.N}](n) < e_{i,j}[\text{sW.T}] \\
    G^{LT}_{i,j}(n) & \text{ if }  e_{i,j}[\text{sW.T}] \leq x_i[\text{sV.N}](n) \leq f_{i,j}[\text{sW.T}], \\
    1 & \text{ if } x_i[\text{sV.N}](n) > f_{i,j}[\text{sW.T}] \\
    \end{array}
    \right.
\end{eqnarray}
with $\mu^{LT}_{i,j}(\cdot)$ the left trapezoidal function, $e_{i,j}[\text{sW.T}$ and $f_{i,j}[\text{sW.T}]$ constants ($e_{i,j}[\text{sW.T}$ < $f_{i,j}[\text{sW.T}]$) and
\begin{equation}
    G^{LT}_{ij}(n) = \frac{x_i[\text{sV.N}](n)- e_{i,j}[\text{W.T}]}{f_{i,j}[\text{W.T}] - e_{i,j}[\text{W.T}]}.
\end{equation} 
Finally, for the triangular membership function is expressed as
\begin{eqnarray}
    \mu^{T}_{i,j}(x_i\text{[sV.N}](n)) =  \left\{ \begin{array}{ll}
    \mu^{LT}_{i,j}(x_i\text{[sV.N}](n)) & \text{if } x_i[\text{sV.N}](n) < m_{i,j}[\text{sW.T}] \\
    \\
    \mu^{RT}_{i,j}(x_i\text{[sV.N}](n)) & \text{if } x_i[\text{sV.N}](n) \geq m_{i,j}[\text{sW.T}] \\
    \end{array}
    \right.,   
\end{eqnarray}
where $m_{i, j}[\text{sW.T}]$ is the triangle center point, that is, $m_{i, j}[\text{sW.T}] = c_{i,j}[\text{sW.T}] = f_{i,j}[\text{sW.T}]$. The values of $\text{W}$ and $\text{T}$ will set the resolution of the activation functions. In the implementation proposed in this work, the value of $\text{W}$ is always expressed as $\text{W}=2 \times \text{T}+1$. The use of non-linear pertinence functions can be accomplished by applying Lookup Tables (LUTs) in the implementation.

Although this implementation uses only two inputs ($x_0[\text{sV.N}](n)$ and $x_1[\text{sV.N}](n)$) and seven membership functions for each input, this can be easily extended for more inputs and functions, since the entire implementation is performed in parallel.

\subsubsection{Operation Module (OM)}\label{sec4}

\begin{figure}[ht]
    \centering
    \includegraphics[width=0.55\columnwidth]{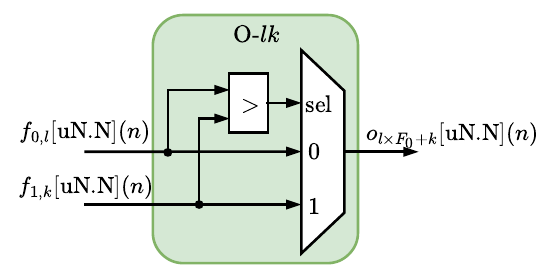}
    \caption{Arquitecture of the module $\text{O-}lk$ associated with the operation between the fuzzyfied signal from the $l$-th membership function from input $0$, $f_{0, l}[\text{nN.N}](n)$, with the $k$-th membership function from input $1$, $f_{1, k} [\text{uN.N}](n)$ (see Equation \ref{EqOFM}).}
    \label{Olk}
\end{figure}

The $F_0 + F_1$ outputs from the MFM module are passed to the OM module that performs all operations relative to the $F_0F_1$ rules, as described in Equation \ref{EqOM} on Section \ref{sec2}. Figure \ref{Olk} details the hardware structure of one of the $F_0F_1$ operating modules, here called $\text{O-}lk$, which performs the minimum operation ("AND" connector) between the $l$-th membership function from input $0$, $f_{0, l}[\text{nN.N}](n)$, with the $k$-th membership function from input $1$, $f_{1, k} [\text{uN.N}](n)$ (see Equation \ref{EqOFM}).

\subsubsection{Output Function Module (OFM)}

\begin{figure}[ht]
    \centering
    \includegraphics[width=0.95\textwidth]{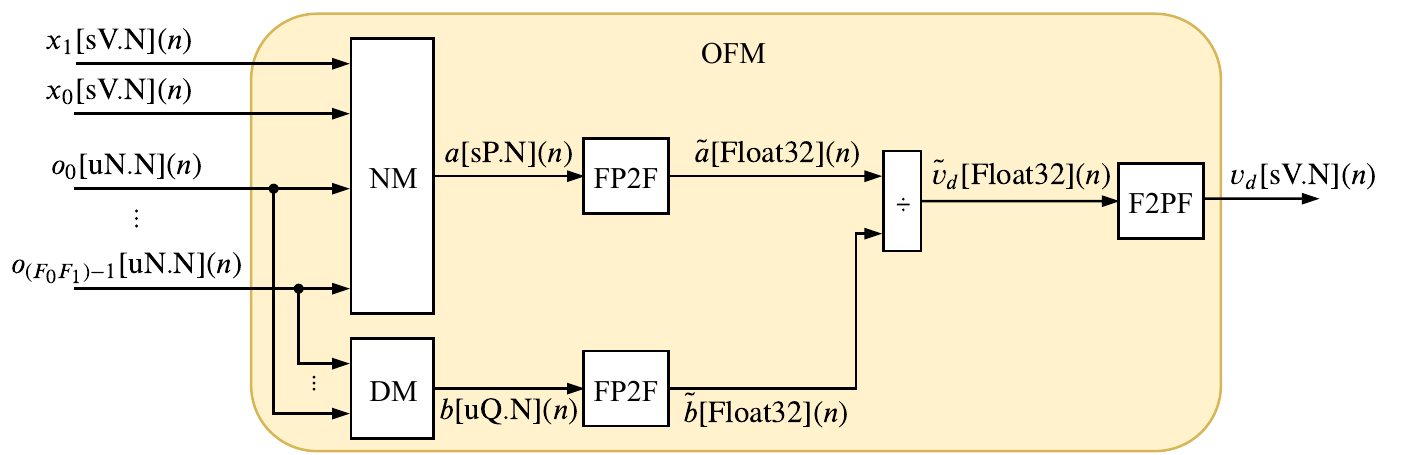}
    \caption{Hardware architecture of the OFM.}
    \label{OFM}
\end{figure}

The OFM, illustrated in Figure \ref{OFM}, performs the generation of the TS-FIMM output variable during the step called defuzzification. This step essentially corresponds to the implementation of the Equation \ref{EqOFM} presented in Section \ref{sec2}. The blocks called NM and DM perform the numerator and denominator operations presented in Equation \ref{EqOFM}, respectively. 

\begin{figure}[ht]
    \centering
    \includegraphics[width=1\textwidth]{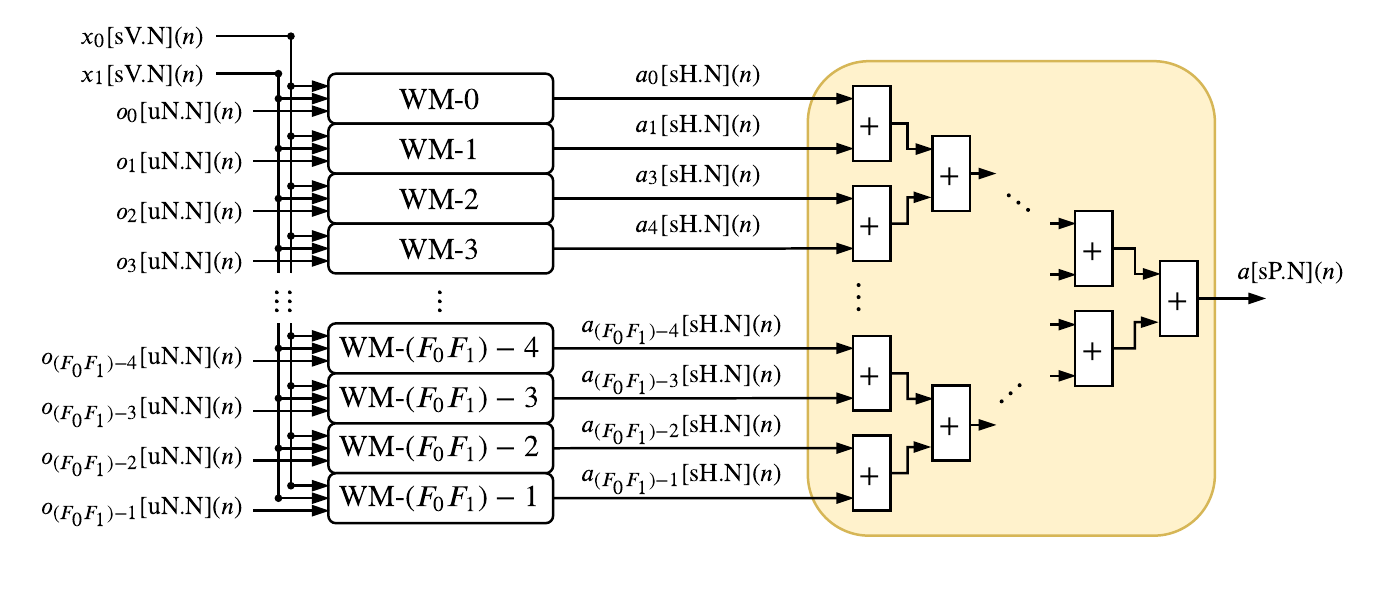}
    \caption{Hardware architecture of the NM.}
    \label{NMOFM}
\end{figure}

\begin{figure}[ht]
    \centering
    \includegraphics[width=0.65\textwidth]{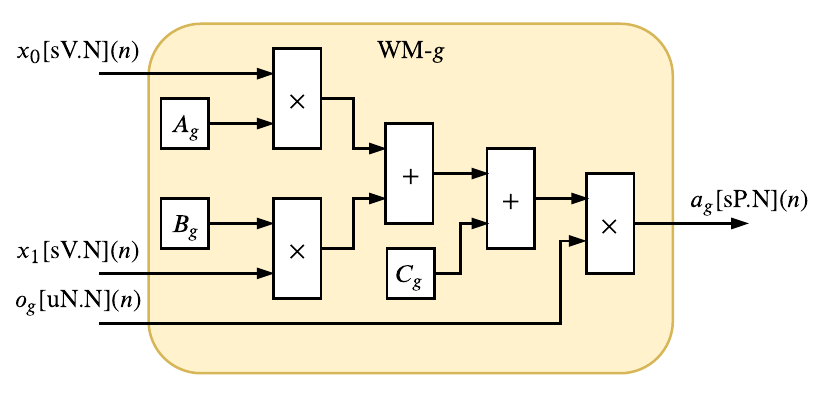}
    \caption{Hardware architecture of the WM-$g$.}
    \label{NMPOFM}
\end{figure}

Figures \ref{NMOFM} and \ref{NMPOFM} show the hardware implementation of the NM. The NM is composed of the $F_0F_1$ hardware components called $\text{WM-}g$ and an adder tree structure. Each $g$-th $\text{WM-}g$, detailed in Figure \ref{NMPOFM}, is a parallel hardware implementation of the variable $a_g$ presented in Equation \ref{EqOFM}. The $F_0F_1$ \text{WMs} hardware components are also implemented in parallel and they generated $F_0F_1$ signals $a_g[\text{sH.N}](n)$ in each $n$-th time instant. Since $-1<x_0[\text{V.N}] (n)<1$, $-1<x_1[\text{V.N}] (n)<1$, $0<o_g[\text{uN.N}] (n)<1$, $-1<A_g<1$, $-1 <B_g<1$ and $-1<C_g<1$ for $g=0,\dots,F_0F_1$ the variable $H$ can be expressed as $H= N + 3$.

\begin{figure}[ht]
    \centering
    \includegraphics[width=0.8\textwidth]{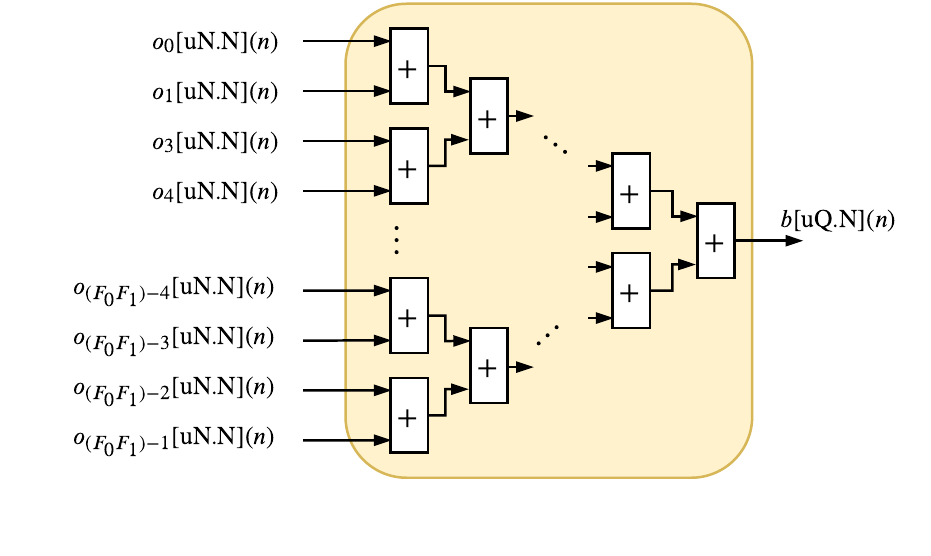}
    \caption{Hardware architecture of the DM.}
    \label{DMOFM}
\end{figure}

The adder tree structure, illustrated in Figure \ref{NMOFM}, has a depth expressed as $\log_2(\lceil F_0F_1 \rceil)$ thus the output signal $a(n)$ (see Equation  \ref{EqOFM}) can be performed as $a[\text{sP.N}](n)$ where
\begin{equation}
    P = H + \log_2(\lceil F_0F_1 \rceil).
\end{equation}
The DM, presented in Figure \ref{DMOFM}, is characterized with an adder tree structure with depth also expressed as $\log_2(\lceil F_0F_1 \rceil)$. The output signal of DM can be expressed as $b[\text{sQ.N}](n)$ where
\begin{equation}
    Q = N + \log_2(\lceil F_0F_1 \rceil) + 1.
\end{equation}

For the division calculation, the output signals, in fixed-point, of the NM and DM modules ($a[\text{sP.N}](n)$ and $b[\text{sQ.N}](n)$ are transformed to a 32-bit floating-point (IEEE754) standard by the Fixed-point to Float (FP2F) module ($\tilde{a}[\text{Float32}](n)$ e $\tilde{b}[\text{Float32}](n)$) and after division the TS-FIMM output is converted back into fixed-point by the Float to Fixed-point (F2FP) module. 

Since the TS-FIMM inputs and the values of $A_g$, $B_g$ and $C_g$ are between $-1$ and $1$, it can be guaranteed, from Equation \ref{EqOFM}, that the output , $v_d[\text{sV.N}](n)$, continue normalized between $-1 $ and $1$. Thus, one can use the same input resolution, that is, $\text{N} $ for the fractional part and $V=N+1$ for the integer part, as shown in Figure \ref{OFM}.

\subsection{Integration Module (IM)}

\begin{figure}[ht]
    \centering
    \includegraphics[width=0.6\textwidth]{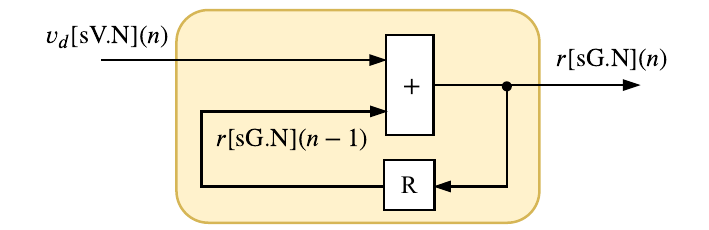}
    \caption{Hardware architecture of the IM.}
    \label{IMHardware}
\end{figure}

The IM, shown in Figure \ref{IMHardware}, implements the Equation \ref{EqInt} presented in Section \ref{sec2}. This module is the last step on the Fuzzy-PI hardware and it is composed of the accumulator with a saturation. The output signal, $r(n)$, is expressed as $r[\text{sG.N}](n)$ where
\begin{equation}
G = N + \log_2(\lceil v_{\text{max}} - v_{\text{min}} \rceil) + 1.
\end{equation}

\section{Synthesis Results}

The synthesis results were obtained to Fuzzy-PI controller (see Figure \ref{GeralHardware}) and also to specific modules TS-FIMM-OS (see Figure \ref{TSFIMHardwareP1}) and TS-FIMM-P (see Figure \ref{TSFIMHardwareP2}). The separate synthesis of the TS-FIMM allows to analysis of the Fuzzy inference algorithm core in the complete hardware proposal. All synthesis results used an FPGA Xilinx Virtex 6 xc6vlx240t-1ff1156 and that has $301$,$440$ registers, $150$,$720$ logical cells to be used as LUTs and $768$ multipliers.

\subsection{Synthesis Results - TS-FIMM Hardware}

Tables \ref{Tab1Results} and \ref{Tab2Results} present the synthesis results related to hardware occupancy and the maximum throughput, $R_s = 1 / t_s$, in Mega samples per second (Msps) of the system for several values of $N$ and $T$. Tables ref{Tab1sults} and \ref{Tab2Results} show the synthesis results associated with TS-FIMM-OS and TS-FIMM-P, respectively. The columns, $\text{NR}$, $\text{NLUT}$ and $\textnormal{NMULT}$ represent the number of registers, logic cells used as LUTs and multipliers in the hardware implemented in the FPGA, respectively. The $\text{PNR}$, $\text{PNLUT}$, and $\textnormal{NMULT}$ columns represent the percentage relative to the total FPGA resources.

\begin{table}[ht]
	\begin{center}
		\caption{Synthesis results (hardware requirement and time) associated with TS-FIMM-OS hardware.}
		\begin{tabular}{cccccccccc}
			\hline
			$\text{N}$ & $\text{T}$ &   $\text{NR}$ & $\text{PR}$ & $\text{NLUT}$  & $\text{PLUT}$ & $\text{NMULT}$ & $\text{PNMULT}$ &$t_s$ ($\text{ns}$) & $R_s$ ($\text{Msps}$) \\
			\hline
			\multirow{4}{*}{$8$} & $4$ & \multirow{4}{*}{$217$} & \multirow{4}{*}{$\approx 0.07\%$} & $6339$ & $\approx 4.21 \%$ & \multirow{4}{*}{$49$} & \multirow{4}{*}{$\approx 6.38\%$} & $79.72$ &  $12.54$ \\
			 & $6$ &  & & $6381$ & $\approx 4.23 \%$ &  &  & $80.95$ &  $12.35$ \\
			 & $8$ &  &  & $6452$ &$\approx 4.28 \%$  &  &  & $81.96$ & $12.20$ \\	
             & $10$ &  &  & $6598$ & $\approx 4.38 \%$ &  &  & $83.76$ &  $11.94$ \\		
			\hline
			\multirow{4}{*}{$10$} & $4$ & \multirow{4}{*}{$259$} & \multirow{4}{*}{$\approx 0.09\%$} & $6772$ & $\approx 4.49 \%$ & \multirow{4}{*}{$49$} & \multirow{4}{*}{$\approx 6.38\%$} & $84.18$ &  $11.88$  \\
		    & $6$ & &  & $6904$ & $\approx 4.58 \%$ &  &  & $82.70$ &  $12.09$ \\	
		    & $8$ & &  & $7331$ & $\approx 4.86 \%$ &  &  & $83.94$ &  $11.91$  \\
		    & $10$ &  &  & $7331$ & $\approx 4.86 \%$  &  &  & $83.00$ &  $12.05$  \\
			\hline
			\multirow{4}{*}{$12$} & $4$ &\multirow{4}{*}{$324$} & \multirow{4}{*}{$\approx 0.11\%$} & $7280$ & $\approx 4.83 \%$ & \multirow{4}{*}{$49$} & \multirow{4}{*}{$\approx 6.38\%$} & $82.65$ &  $12.10$ \\
			& $6$ & &  & $7916$ & $\approx 5.25 \%$ &  &  & $83.28$ &  $12.01$  \\
			& $8$ & &  & $7954$ &  $\approx 5.28 \%$ &  &  & $87.02$ &  $11.49$  \\
			& $10$ & &  & $8147$ & $\approx 5.41 \%$  &  &  & $85.99$ &  $11.63$  \\
		    \hline
			\multirow{4}{*}{$14$} &$4$ & \multirow{4}{*}{$384$} & \multirow{4}{*}{$\approx 0.13\%$} & $8761$ & $\approx 5.81 \%$ & \multirow{4}{*}{$49$} & \multirow{4}{*}{$\approx 6.38\%$} & $84.12$ &  $11.89$ \\
			&$6$ &  &  & $8915$ &  $\approx 5.91 \%$ &  &  & $85.08$ &  $11.75$  \\
			&$8$ &  &  & $8999$ &  $\approx 5.97 \%$ &  &  & $86.39$ &  $11.58$  \\
			&$10$ &  &  & $9163$ &  $\approx 6.08 \%$ &  &  & $86.75$ &  $11.53$  \\
			\hline
			\multirow{4}{*}{$16$} & $4$ &\multirow{4}{*}{$428$} & \multirow{4}{*}{$\approx 0.14\%$} & $9816$ & $\approx 6.51 \%$ & \multirow{4}{*}{$49$} & \multirow{4}{*}{$\approx 6.38\%$} & $86.42$ &  $11.54$ \\
			& $6$ & &  & $9990$ & $\approx 6.63 \%$ &  &  & $84.80$ &  $11.79$  \\
			& $8$ & &  & $10072$ & $\approx 6.68 \%$ &  &  & $88.31$ &  $11.32$  \\
			& $10$ & &  & $10252$ & $\approx 6.80 \%$ &  &  & $88.65$ &  $11.28$  \\
			\hline        
		\end{tabular}
		\label{Tab1Results}
	\end{center}
\end{table}

\begin{table}[ht]
	\begin{center}
		\caption{Synthesis results (hardware requirement and time) associated with TS-FIMM-P hardware.}
		\begin{tabular}{cccccccccc}
			\hline
			$\text{N}$ & $\text{T}$ &   $\text{NR}$ & $\text{PR}$ & $\text{NLUT}$  & $\text{PLUT}$ & $\text{NMULT}$ & $\text{PNMULT}$ &$t_s$ ($\text{ns}$) & $R_s$ ($\text{Msps}$) \\
			\hline
			\multirow{4}{*}{$8$} & $4$ & \multirow{4}{*}{$746$} & \multirow{4}{*}{$\approx 0.25\%$} & $5326$ & $\approx 3.53\%$ & \multirow{4}{*}{$49$} & \multirow{4}{*}{$\approx 6.38\%$} & $56.73$ &  $17.62$ \\
			 & $6$ &  & & $5350$ & $\approx 3.55\%$ &  &  & $55.81$ &  $17.92$  \\
			 & $8$ &  &  & $5422$ & $\approx 3.60\%$ &  &  & $56.18$ & $17.80$  \\	
             & $10$ &  &  & $5590$ &  $\approx 3.71\%$ &  &  & $56.97$ &  $17.55$ \\		
			\hline
			\multirow{4}{*}{$10$} & $4$ & \multirow{4}{*}{$917$} & \multirow{4}{*}{$\approx 0.30\%$} & $6093$ & $\approx 4.04\%$ & \multirow{4}{*}{$49$} & \multirow{4}{*}{$\approx 6.38\%$} & $57.21$ &  $17.48$  \\
		    & $6$ & &  & $6141$ & $\approx 4.07\%$ &  &  & $57.88$ &  $17.28$ \\	
		    & $8$ & &  & $6199$ &  $\approx 4.11\%$ &  &  & $57.63$ &  $17.35$  \\
		    & $10$ &  &  & $6317$ &  $\approx 4.19\%$ &  &  & $56.72$ &  $17.63$  \\
			\hline
			\multirow{4}{*}{$12$} & $4$ &\multirow{4}{*}{$1113$} & \multirow{4}{*}{$\approx 0.37\%$} & $6910$ & $\approx 4.58\%$ & \multirow{4}{*}{$49$} & \multirow{4}{*}{$\approx 6.38\%$} & $57.90$ &  $17.27$ \\
			& $6$ & &  & $6982$ & $\approx 4.63\%$ &  &  & $58.22$ &  $17.18$  \\
			& $8$ & &  & $7016$ & $\approx 4.65\%$ &  &  & $58.60$ &  $17.06$  \\
			& $10$ & &  & $7172$ &  $\approx 4.76\%$ &  &  & $56.26$ &  $17.77$  \\
		    \hline
			\multirow{4}{*}{$14$} &$4$ & \multirow{4}{*}{$1301$} & \multirow{4}{*}{$\approx 0.43\%$} & $7799$ & $\approx 5.17\%$ & \multirow{4}{*}{$49$} & \multirow{4}{*}{$\approx 6.38\%$} & $58.60$ &  $17.06$ \\
			&$6$ &  &  & $7823$ & $\approx 5.19\%$ &  &  & $58.22$ &  $17.18$  \\
			&$8$ &  &  & $7905$ &  $\approx 5.24\%$ &  &  & $58.26$ &  $17.16$  \\
			&$10$ &  &  & $8031$ &  $\approx 5.33\%$ &  &  & $60.00$ &  $16.66$  \\
			\hline
			\multirow{4}{*}{$16$} & $4$ &\multirow{4}{*}{$1477$} & \multirow{4}{*}{$\approx 0.49\%$} & $8713$ & $\approx 5.78\%$ & \multirow{4}{*}{$49$} & \multirow{4}{*}{$\approx 6.38\%$} & $59.43$ &  $16.83$ \\
			& $6$ & &  & $8737$ & $\approx 5.80\%$ &  &  & $58.14$ &  $17.20$  \\
			& $8$ & &  & $8819$ &  $\approx 5.85\%$ &  &  & $57.89$ &  $17.27$  \\
			& $10$ & &  & $8955$ & $\approx 5.94\%$  &  &  & $58.90$ &  $16.98$  \\
			\hline   
		\end{tabular}
		\label{Tab2Results}
	\end{center}
\end{table}

Synthesis results show that the hardware proposal for TS-FIMM takes up a small hardware space of less than $1\%$, $\text{PR}$, in registers and less than $7\%$ in LUTs, $ \text{PLUT}$, of the FPGA (see Tables \ref{Tab1Results} and \ref{Tab2Results}). These results enable the use of several TS-FIMM implemented in parallel on FPGA, allowing to accelerate several applications in massive data environments. On the other hand, the low hardware consumption allows the use of TS-FIMM in small FPGAs of low cost and consumption for applications of IoT and M2M. Another important point to be analyzed, still in relation to the synthesis, is the linear behavior of the hardware consumption in relation to the number of bits, unlike the work presented in \cite {Marcelo1}, and this is important, since it makes possible the use systems with higher resolution.

The values of throughput, $R_s$, were very relevant, with values about $11.5 \text{Msps}$ for TS-FIMM-OS and values about $17 \text{Msps}$ for TS-FIMM-P. These values enables its application in various large volume problems for processing as presented in \cite {PaperFPGAFuzzy3} or in problems with fast control requirements such as tactile internet applications \cite{tactile, Tactile1RESRC}. It is also observed that throughput has a linear behavior as a function of the number of bits.

The TS-FIMM-P has a speedup about $1.47 \times$ ($\frac{17 \text{Msps}}{11.5 \text{Msps}}$) regards the TS-FIMM-OS. This speedup was driven by the critical path reduction with the pipeline scheme. However, the pipeline scheme in TS-FIMM-P used about $3.4 \times$ registers ($\text{NR}$) more than TS-FIMM-OS. 

\begin{figure}[ht]
    \centering
    \includegraphics[width=0.7\textwidth]{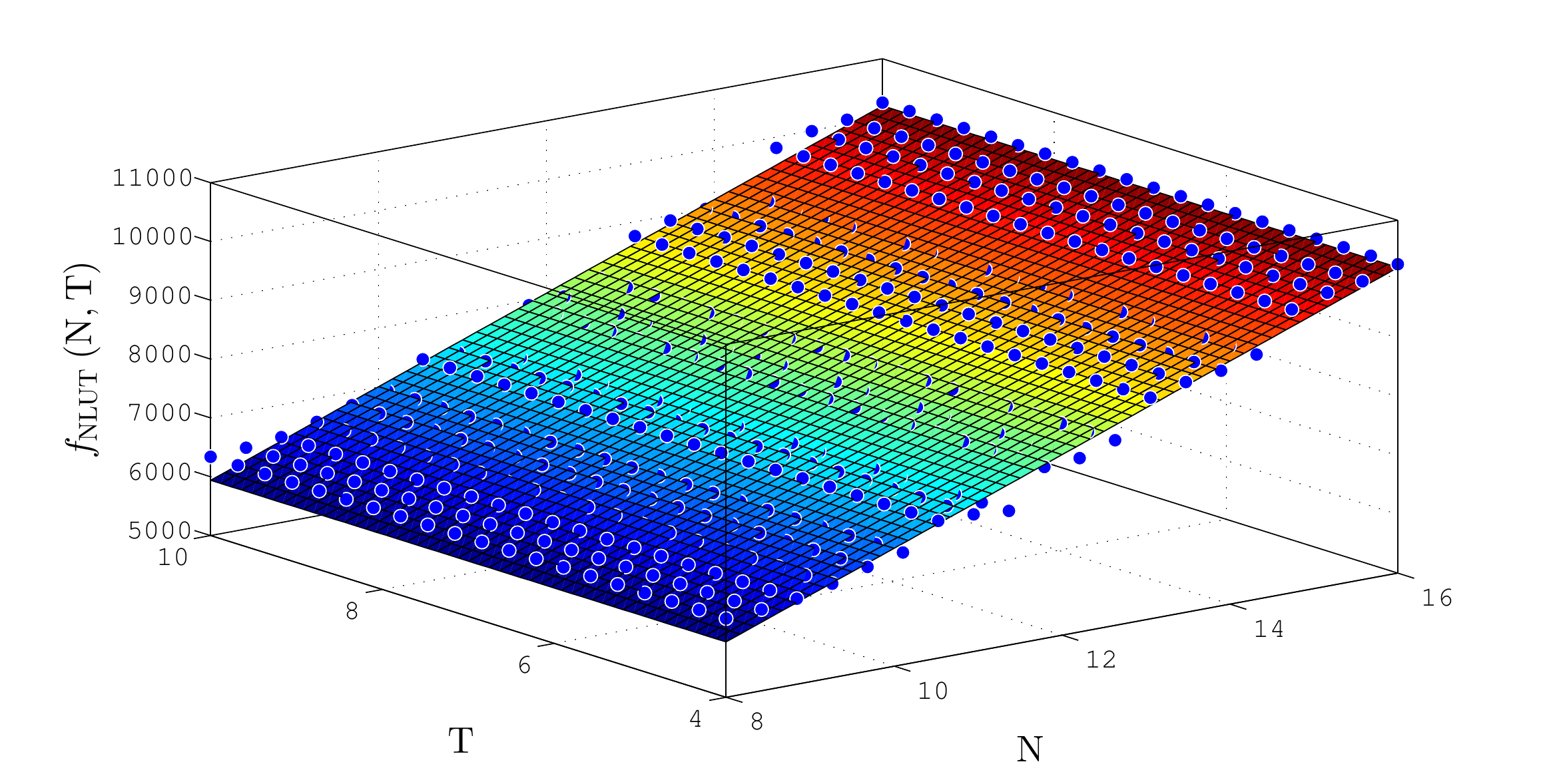}
    \caption{Plane, $f_{\text{NLUT}}\left(\text{N},\text{T} \right)$, found to estimate the number of LUTs in function of the number of bits $\text{N}$ and $\text{T}$ for TS-FIMM-OS.}
    \label{PlanoNLUTOS}
\end{figure}

\begin{figure}[ht]
\centering
\includegraphics[width=0.7\textwidth]{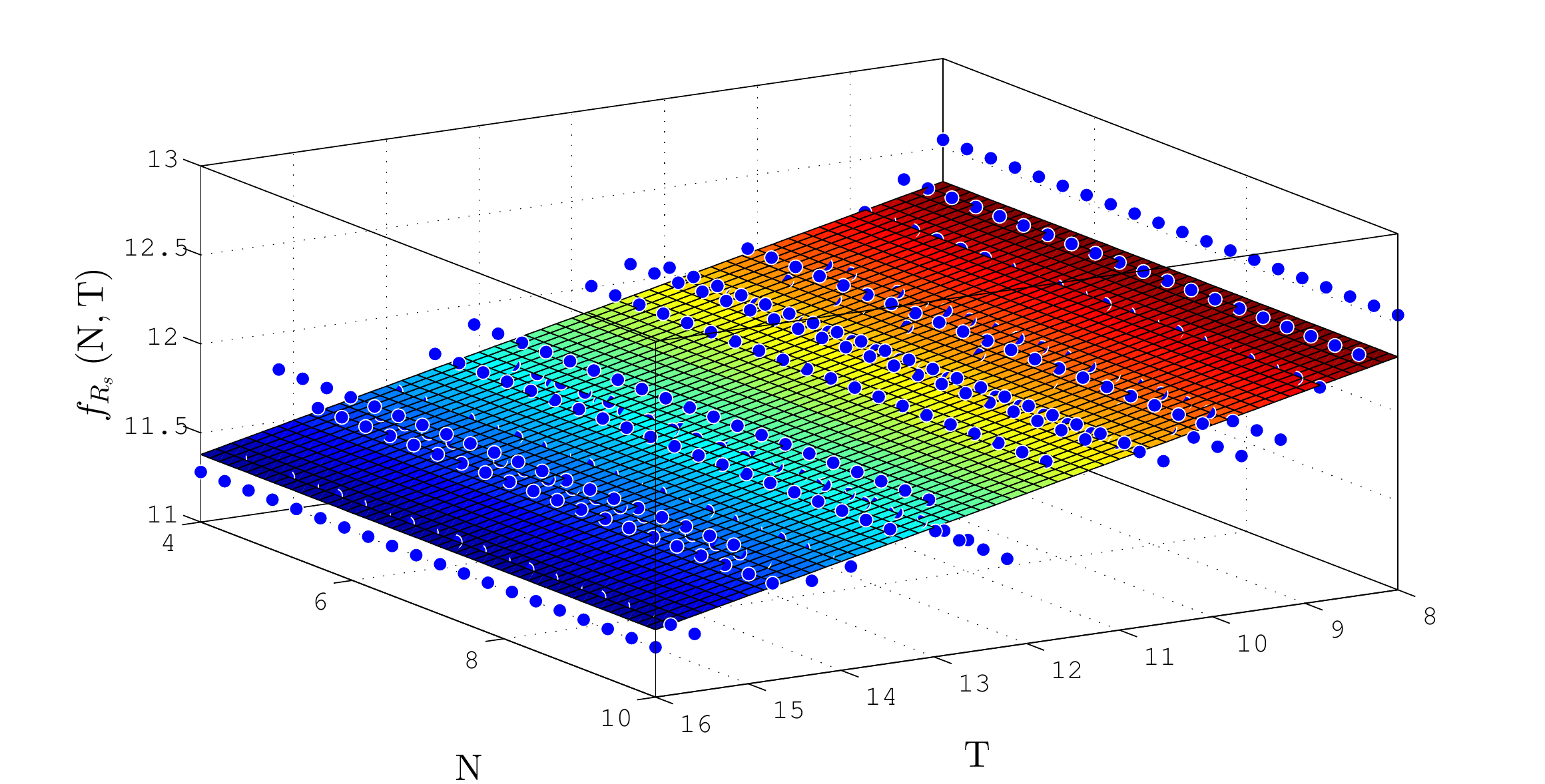}
\caption{Plane, $f_{R_s}\left(\text{N},\text{T} \right)$, found to estimate throughput, $R_s$, for different number of bits $\text{N}$ and $\text{T}$ for TS-FIMM-OS.}
\label{PlanoRsOS}
\end{figure}

The figures \ref{PlanoNLUTOS} and \ref{PlanoRsOS} show the behavior surfaces of the number of LUTs ($\text{NLUT}$) and throughput in function of $\text{N}$ and $ \text{T}$ for TS-FIMM-OS, respectively. For both cases an adjustment was made, through a regression technique, to find the plane that best matches the measured points. For the case of $\text{NLUT} $, the plane, $f_\text{NLUT}\left(\text{N},\text{T} \right)$ expressed by
\begin{equation}
f_{\text{NLUT}}\left(\text{N},\text{T} \right) \approx 1682 +  532.2 \times \text{N} +  6.493\times 10^{-13} \times \text{T},
\end{equation}
with a $R^2=0.9766$. For throughput in $\text{Msps}$ was found a plane, $f_{R_s}\left(\text{N},\text{T} \right)$, characterized as
\begin{equation}
f_{R_s}\left(\text{N},\text{T} \right) \approx 13.24 -  0.1163 \times \text{N} +  3.414\times 10^{-16} \times \text{T},
\end{equation}
with $R^2=0.7521$. 

\begin{figure}[ht]
    \centering
    \includegraphics[width=0.7\textwidth]{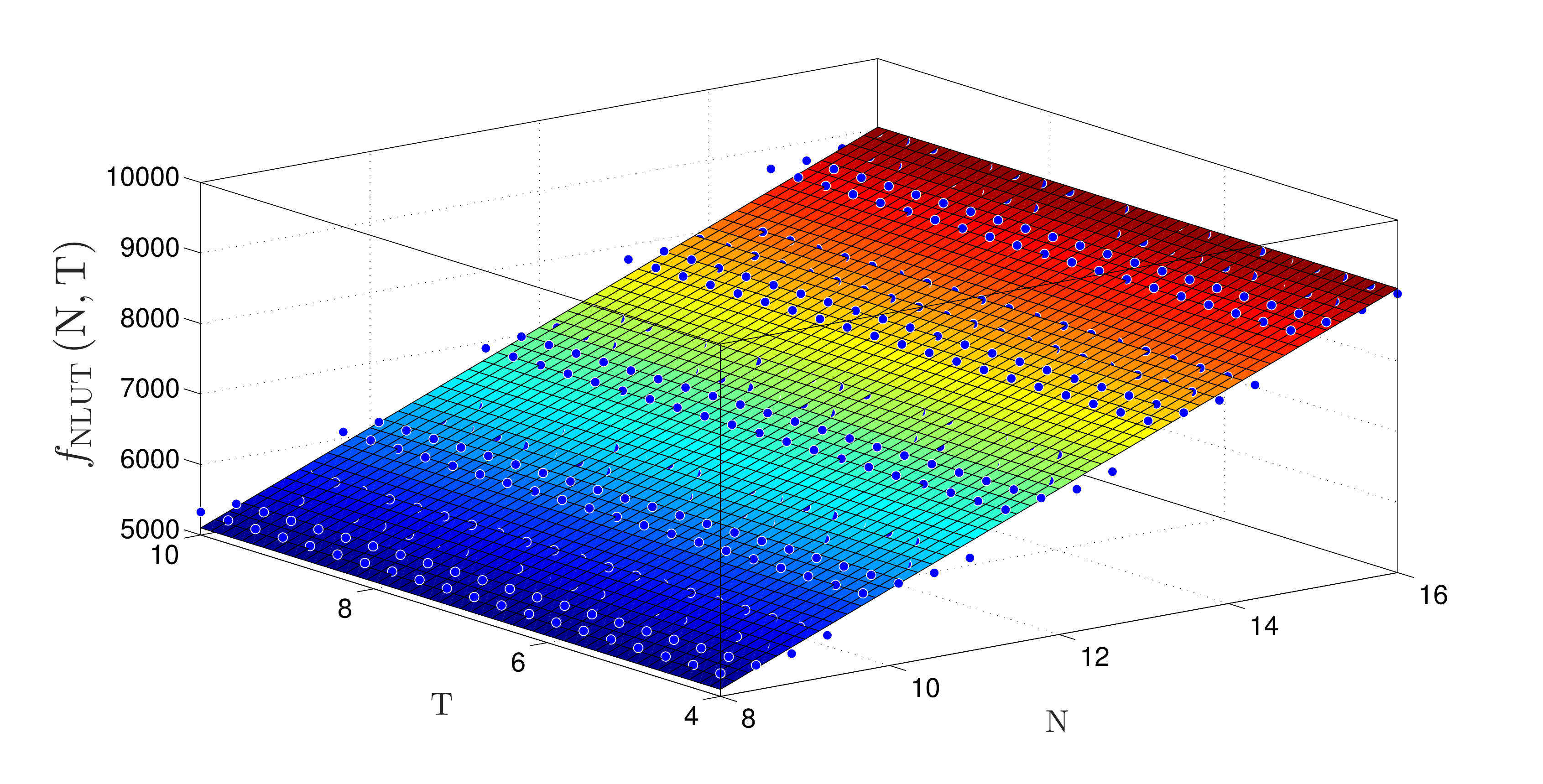}
    \caption{Plane, $f_{\text{NLUT}}\left(\text{N},\text{T} \right)$, found to estimate the number of LUTs in function of the number of bits $\text{N}$ and $\text{T}$ for TS-FIMM-P.}
    \label{PlanoNLUTP}
\end{figure}

\begin{figure}[ht]
\centering
\includegraphics[width=0.7\textwidth]{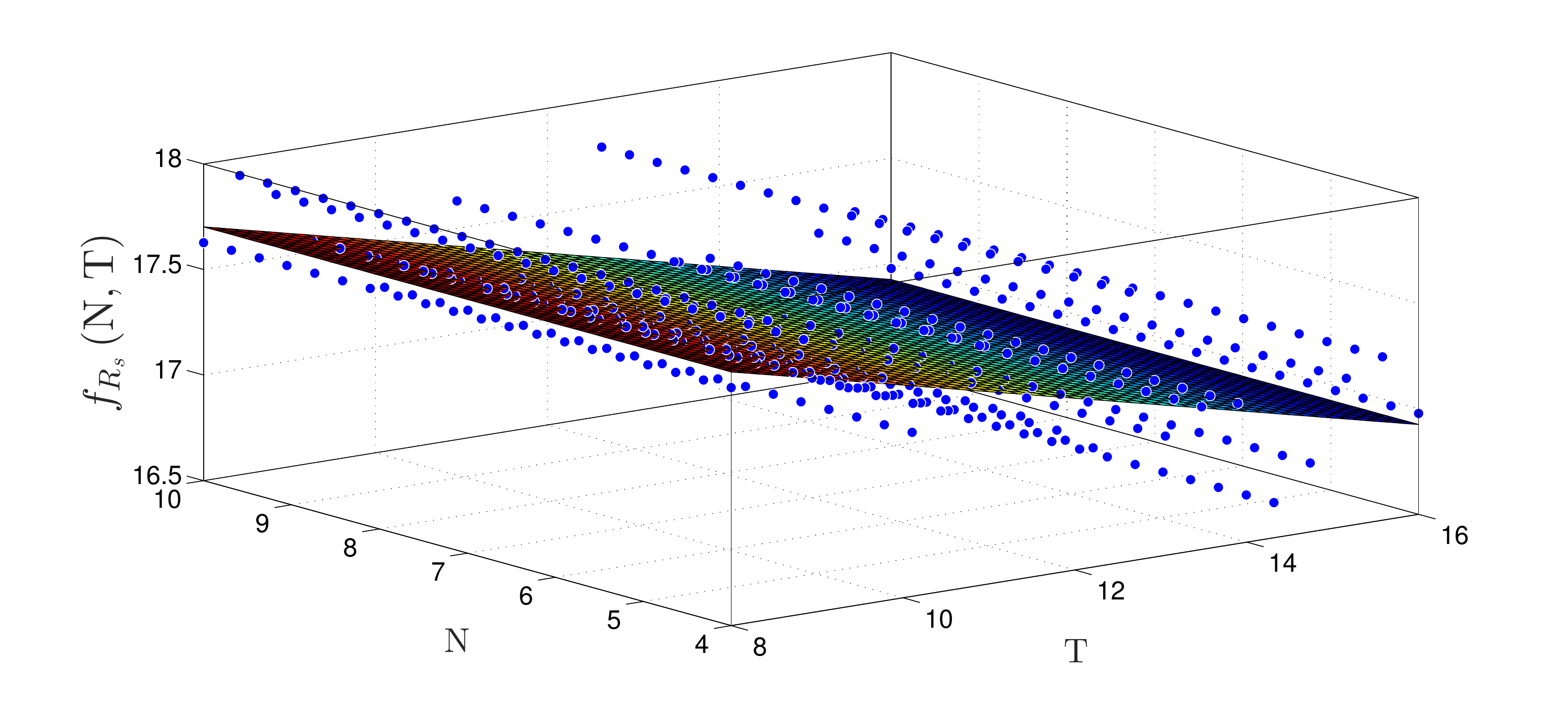}
\caption{Plane, $f_{R_s}\left(\text{N},\text{T} \right)$, found to estimate throughput, $R_s$, for different number of bits $\text{N}$ and $\text{T}$ for TS-FIMM-P.}
\label{PlanoRsP}
\end{figure}

The behavior surfaces of the number of LUTs ($\text{NLUT}$) and throughput in function of $\text{N}$ and $ \text{T}$ for TS-FIMM-P are presented in Figures \ref{PlanoNLUTP} and \ref{PlanoRsP}, respectively. For the case of $\text{NLUT} $, the plane, $f_\text{NLUT}\left(\text{N},\text{T} \right)$ expressed by
\begin{equation}
f_{\text{NLUT}}\left(\text{N},\text{T} \right) \approx 1171 +  491.1 \times \text{N} +  4.245\times 10^{-13} \times \text{T},
\end{equation}
with a $R^2=0.9838$. For throughput in $\text{Msps}$ was found a plane, $f_{R_s}\left(\text{N},\text{T} \right)$, characterized as
\begin{equation}
f_{R_s}\left(\text{N},\text{T} \right) \approx 18.48 -  0.09704 \times \text{N} - 5.365\times 10^{-16} \times \text{T},
\end{equation}
with $R^2=0.5366$. 

\subsection{Synthesis Results - Fuzzy-PI Controller Hardware}

Tables \ref{Tab3Results} and \ref{Tab4Results} present the synthesis results related to hardware occupancy and throughput, $R_s$ for the Fuzzy-PI controller hardware (see Figure \ref{GeralHardware}) . The results are presented for several values of $N$ and $T=10$. 

\begin{table}[ht]
	\begin{center}
		\caption{Synthesis results (hardware requirement and time) associated with Fuzzy-PI controller hardware with TS-FIMM-OS.}
		\begin{tabular}{ccccccccc}
			\hline
			$\text{N}$ &  $\text{NR}$ & $\text{PR}$ & $\text{NLUT}$  & $\text{PLUT}$ & $\text{NMULT}$ & $\text{PNMULT}$ &$t_s$ ($\text{ns}$) & $R_s$ ($\text{Msps}$) \\
			\hline
			$8$ & $261$ & $\approx 0.09\%$ & $6834$ & $\approx 4.53\%$ & $49$ & $\approx 6.38\%$ & $92.87$ &  $10.77$ \\
			$10$ & $307$ & $\approx 0.10\%$ & $7331$ & $\approx 4.86\%$ & $49$ & $\approx 6.38\%$ & $98.44$ &  $10.16$ \\
			$12$ & $375$ & $\approx 0.12\%$ & $8409$ & $\approx 5.58\%$ & $49$ & $\approx 6.38\%$ & $98.68$ &  $10.13$ \\
			$14$ & $438$ & $\approx 0.15\%$ & $9460$ & $\approx 6.28\%$ & $49$ & $\approx 6.38\%$ & $99.98$ &  $10.00$ \\
			$16$ & $488$ & $\approx 0.16\%$ & $10595$ & $\approx 7.03\%$ & $49$ & $\approx 6.38\%$ & $104.31$ &  $9.59$ \\
			\hline   
		\end{tabular}
		\label{Tab3Results}
	\end{center}
\end{table}

\begin{table}[ht]
	\begin{center}
		\caption{Synthesis results (hardware requirement and time) associated with Fuzzy-PI controller hardware with TS-FIMM-P.}
		\begin{tabular}{ccccccccc}
			\hline
			$\text{N}$ &  $\text{NR}$ & $\text{PR}$ & $\text{NLUT}$  & $\text{PLUT}$ & $\text{NMULT}$ & $\text{PNMULT}$ &$t_s$ ($\text{ns}$) & $R_s$ ($\text{Msps}$) \\
			\hline
			$8$ & $790$ & $\approx 0.26\%$ & $5826$ & $\approx 3.87\%$ & $49$ & $\approx 6.38\%$ & $66.08$ &  $15.13$ \\
			$10$ & $965$ & $\approx 0.32\%$ & $6317$ & $\approx 4.19\%$ & $49$ & $\approx 6.38\%$ & $72.16$ &  $13.86$ \\
			$12$ & $1164$ & $\approx 0.39\%$ & $7434$ & $\approx 4.93\%$ & $49$ & $\approx 6.38\%$ & $68.95$ &  $14.50$ \\
			$14$ & $1355$ & $\approx 0.45\%$ & $8328$ & $\approx 5.53\%$ & $49$ & $\approx 6.38\%$ & $73.23$ &  $13.66$ \\
			$16$ & $1537$ & $\approx 0.51\%$ & $9298$ & $\approx 6.17\%$ & $49$ & $\approx 6.38\%$ & $74.56$ &  $13.41$ \\
			\hline   
		\end{tabular}
		\label{Tab4Results}
	\end{center}
\end{table}

Synthesis results, drawn on Table \ref{Tab3Results} and \ref{Tab4Results}, show that the proposed implementation requires a small fraction of hardware space, less than $1\% $, $\text{PR}$, in registers and less than $8\%$ in LUTs, $ \text{PLUT}$, of the FPGA. In addition, it is possible to see the numbers of embedded multipliers, $\text{PNMULT}$, remained below $7\%$. This occupation enables the use of several Fuzzy-PI controllers in parallel in the same FPGA hardware and this allows various controls systems running in parallel on industrial applications. The low size implementation also allows the use in low cost and power consumption IoT and M2M applications. Regarding throughput, $R_s$, the results obtained were highly relevant, with values between $15.33$, and $13.41\, \text{Msps}$. Which enables its application in several problems with large data volume for processing as presented in \cite{PaperFPGAFuzzy3} or in problems with fast control requirements such as tactile internet applications \cite{Tactile1RESRC}.  

\section{Validation Results}

\subsection{Validation Results - TS-FIMM Hardware}

\begin{figure}[ht]
\centering
\includegraphics[width=.65\columnwidth]{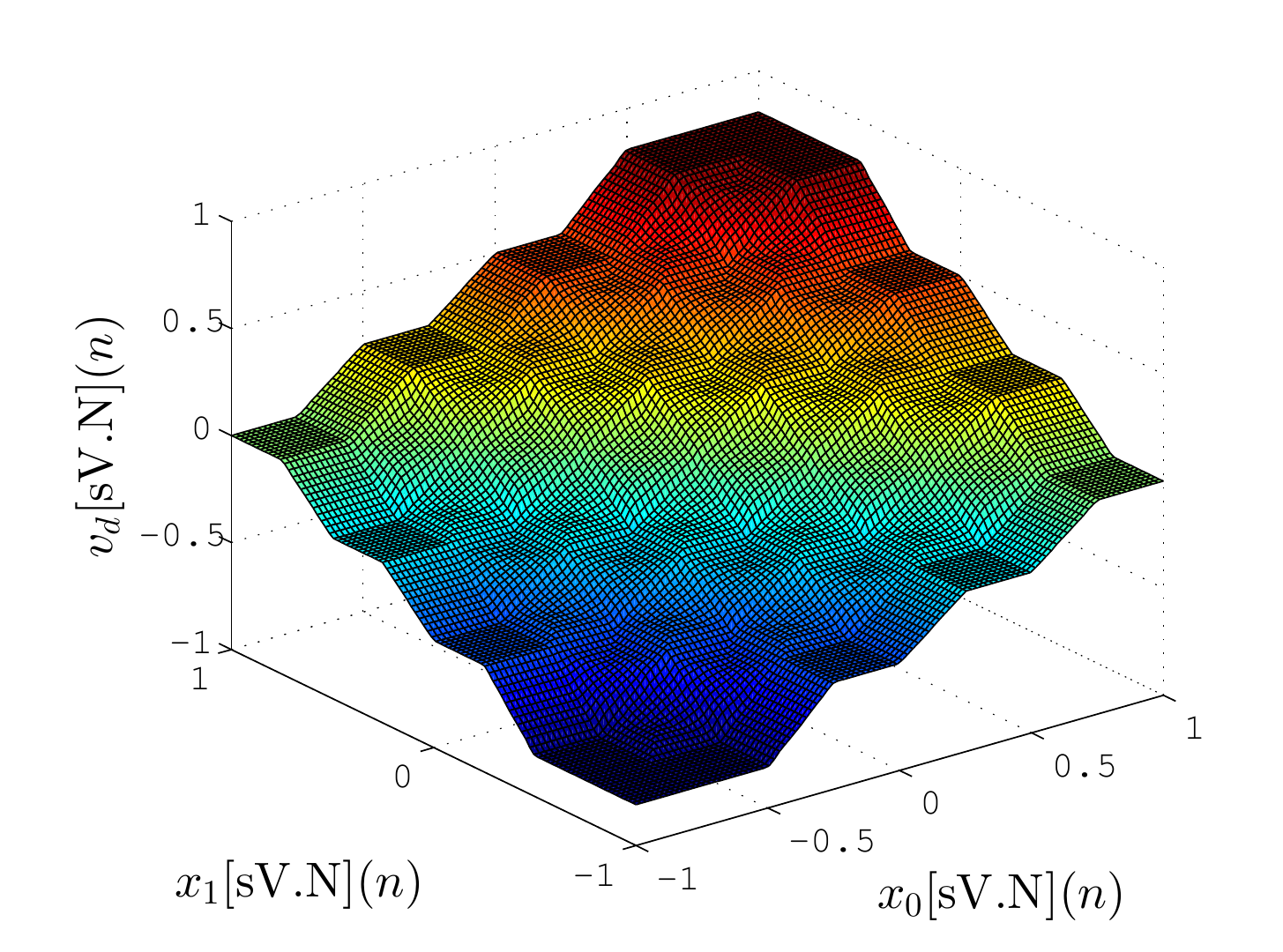}
\caption{Mapping between input and output from TS-FIMM hardware using fixed-point with $\text{N}=8$, $\text{V}=9$ and $\text{T}=4$.}
\label{SurfMappingTS_FIMM_Hardware}
\end{figure}

\begin{figure}[ht]
\centering
\includegraphics[width=.65\columnwidth]{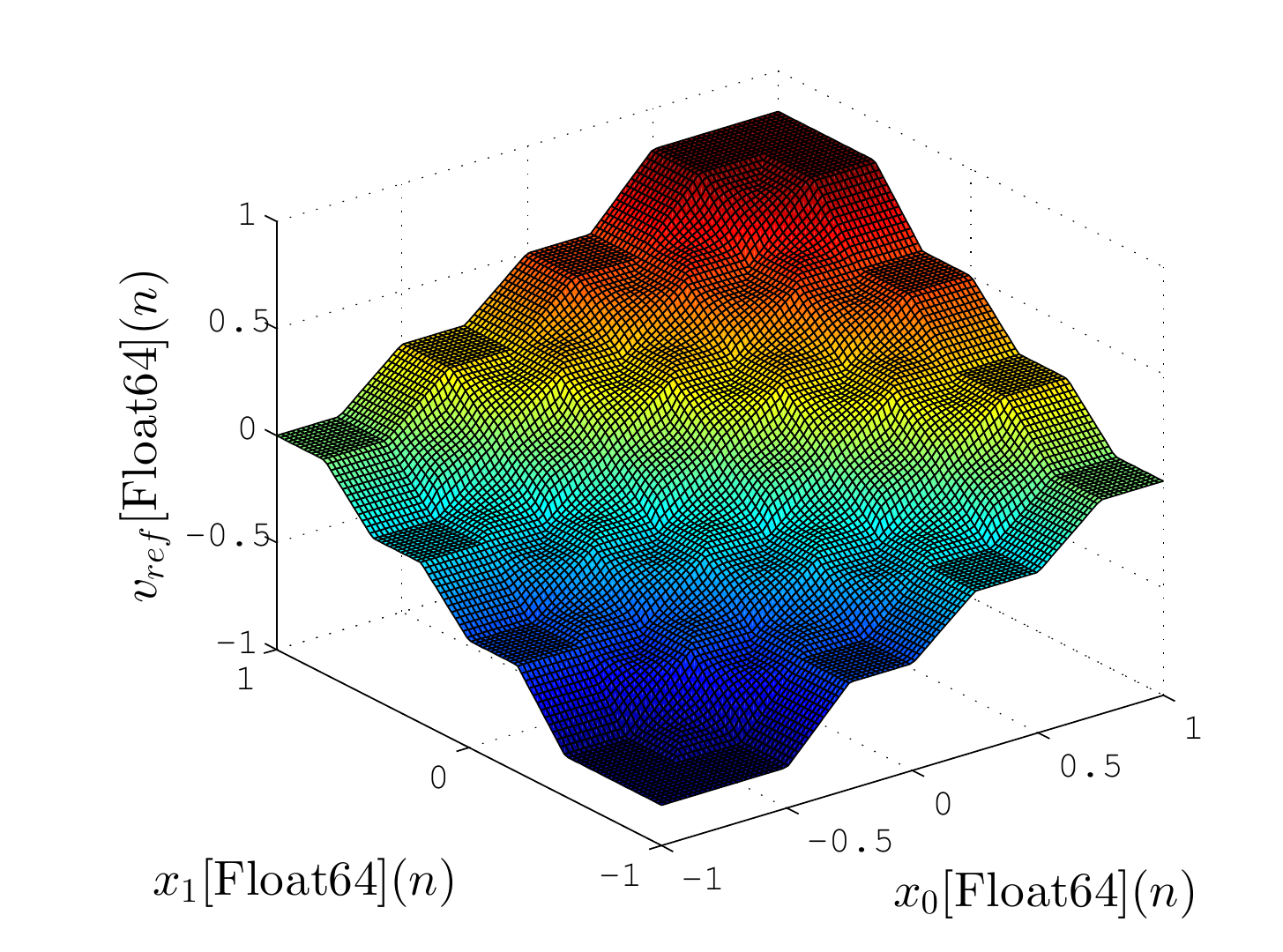}
\caption{Mapping between input and ouput from TS-FIMM generated by Matlab Fuzzy Logic Toolbox using double format.}
\label{SurfMappingTS_FIMM_Double}
\end{figure}

The Figures \ref{SurfMappingTS_FIMM_Hardware} and \ref{SurfMappingTS_FIMM_Double} show the mapping between input ($x_0(n)$ and $x_1(n)$) and output $v_d(n)$ for proposed hardware and a reference implementation with Fuzzy Matlab Toolbox (License number 1080073) \cite{MatlabFuzzyToolbox}, respectively. The Matlab implementation, shown in Figure \ref{SurfMappingTS_FIMM_Double}, uses floating-point format with $64$ bits (double precision) while in Figure \ref{SurfMappingTS_FIMM_Hardware} the proposed hardware-generated mapping is presented using lower resolution synthesized ($N=8$, $V=9$ and $T=4$). These figures are able to present a qualitative representation of the proposed implementation, in which the obtained results are quite similar to those expected.

The Table \ref{TabMSEResults1} shows the mean square error (MSE) between the Fuzzy Matlab Toolbox and the proposed hardware implementation for several cases $N$ and $T$. For the experiment, the calculation of $MSE$ is expressed as
\begin{equation}\label{MSEEq}
MSE = \frac{1}{Z}\sum_{n=0}^{Z-1} \left(v_{ref}[\text{Float64}](n) - v_d[\text{sV.N}](n)\right)^2,
\end{equation}
where $Z$ represents the number of tested points that corresponded to $10000$ points spread evenly within the limits of the input values ($-1$ and $1$). The Figures \ref{SurfMappingTS_FIMM_Hardware} and \ref{SurfMappingTS_FIMM_Double} were generated with these points. 

\begin{table}[ht]
	\small
	\begin{center}
		\caption{Mean square error (MSE) between the Fuzzy Matlab Toolbox and the proposed hardware implementation for several cases $N$ and $T$.}
		\begin{tabular}{ccc}
			\hline
			$\text{N}$ & $\text{T}$ &  $MSE$ (see Equation \ref{MSEEq})\\
			\hline
			\multirow{4}{*}{$8$} & $4$  &  \multirow{4}{*}{$2.4 \times 10^{-6}$}\\
			& $6$ &    \\
			& $8$ &     \\	
			& $10$ &     \\		
			\hline
			\multirow{4}{*}{$10$} & $4$  & \multirow{4}{*}{$1.3 \times 10^{-7}$}\\
			& $6$ &   \\
			& $8$ &    \\
			& $10$ &    \\
			\hline
			\multirow{4}{*}{$12$} & $4$   & \multirow{4}{*}{$7.2 \times 10^{-9}$}\\
			& $6$ &   \\
			& $8$ &   \\
			& $10$ &   \\
			\hline
			\multirow{4}{*}{$14$} &$4$  & \multirow{4}{*}{$4.9 \times 10^{-10}$}\\
			&$6$ &   \\
			&$8$ &  \\
			&$10$ &   \\
			\hline
			\multirow{4}{*}{$16$} & $4$  & \multirow{4}{*}{$2.7 \times 10^{-11}$}\\
			& $6$ &    \\
			& $8$ &   \\
			& $10$ & \\
			\hline        
		\end{tabular}
		\label{TabMSEResults1}
	\end{center}
\end{table}

The results obtained in relation to $MSE$ were also quite significant, showing that the TS-FIMM hardware has a response quite similar to the implementation with $64$ bits even for a fixed-point resolution of $8$ bits ($ MSE = 2$,$395 \times 10^{ -6}$). Another interesting fact was related to the values of $T$ that did not significantly influence the $MSE$ value for the pertinence functions used (see Figure \ref{fig:MFIn1}) in the project. It is important to note that the implementation of TS-FIMM hardware with few bits leads to smaller hardware, low-power consumption or high-throughput values.

\subsection{Validation Results - Fuzzy-PI Controller Hardware}

\begin{figure}[ht]
	\begin{center}
		\includegraphics[width=.5\textwidth]{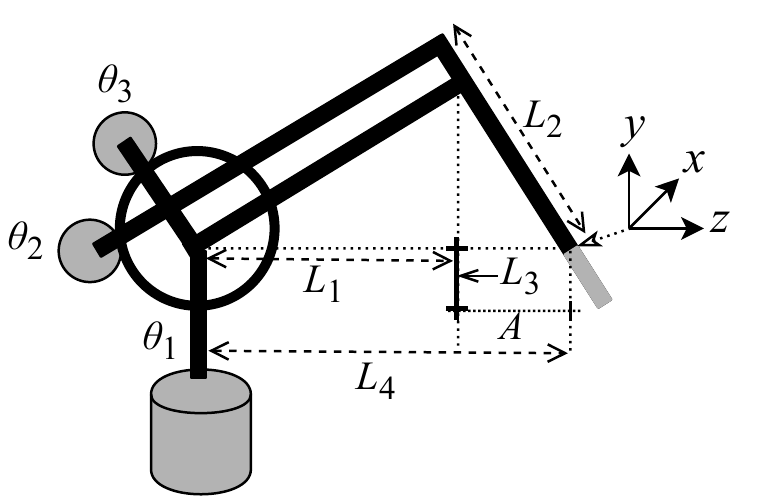}
		\caption{Structure of 3-DOF Phantom Omni robotic manipulator.} 
		\label{RoboticManipulatorModel}
	\end{center}
\end{figure}

In order to validate the results of the Fuzzy-PI controller in hardware, bit-precision simulation tests were performed with a non-linear dynamic system characterized by a robotic manipulator system called the Phantom Omni \cite{4097836, 6318365, 5341989, 7813108}. The Phantom Omni is a 6-DOF (Degree Of Freedom) manipulator, with rotational joints. The first three joints are actuated, while the last three joints are non-actuated \cite{}. As illustrated in Figure \ref{RoboticManipulatorModel}, the device can be modeled as 3-DOF robotic manipulator with two segments $L_1$ and $L_2$. The segments are interconnected by three rotary joints angles $\theta_1$, $\theta_2$ and $\theta_3$. The Phantom Omni has been widely used in literature, as presented in \cite{4097836, 6318365, 5341989}. Simulations used $L_1 = 0.135 \, \text{mm}$, $L2 = L1$, $L3=0.025 \, \text{mm}$ and $L4 = L1+A$ where $A=0.035 \, \text{mm}$ as described in \cite{5341989}.

Non-linear, second order, ordinary differential equation used to describe the dynamics of the Phantom Omni can be expressed as
\begin{equation}
\mathbf{M}\left(\boldsymbol{\theta}(t)\right)\boldsymbol{\ddot{\theta}}(t)+\mathbf{C}\left(\boldsymbol{\theta}(t),\boldsymbol{\dot{\theta}}(t)\right)\boldsymbol{\dot{\theta}}(t)+\mathbf{g}\left(\boldsymbol{\theta}(t)\right) - \mathbf{f}\left(\boldsymbol{\dot{\theta}}(t)\right)= \boldsymbol{\tau}(t)
\end{equation}
where $\boldsymbol{\theta}(t)$ is the vector of joints expressed as
\begin{equation}
\boldsymbol{\theta}(t) = \left[\begin{array}{ccc}
\theta_1(t) & \theta_2(t) & \theta_3(t)
\end{array} \right]^T \in \mathbb{R}^{3 \times 1},
\end{equation}
$\boldsymbol{\tau}$ is the vector of torques acting expressed as
\begin{equation}
\boldsymbol{\tau}(t) = \left[\begin{array}{ccc}
\tau_1(t) & \tau_2(t) & \tau_3(t)
\end{array} \right]^T \in \mathbb{R}^{3 \times 1},
\end{equation}
$\mathbf{M}\left(\boldsymbol{\theta}(t)\right) \in \mathbb{R}^{3 \times 3}$ is the inertia matrix, $\mathbf{C}\left(\boldsymbol{\theta}(t),\boldsymbol{\dot{\theta}}(t)\right) \in \mathbb{R}^{3 \times 3}$ is the Coriolis and centrifugal forces matrix, $\mathbf{g}\left(\boldsymbol{\theta}(t)\right) \in \mathbb{R}^{3 \times 1}$ represents the gravity force acting on the joints, $\boldsymbol{\theta}(t)$, and the $\mathbf{f}\left(\boldsymbol{\dot{\theta}}(t)\right)$ is the friction force on the joints, $\boldsymbol{\theta}(t)$ \cite{4097836, 6318365, 5341989, 7813108}.

\begin{figure}[ht]
	\begin{center}
		\includegraphics[width=.95\textwidth]{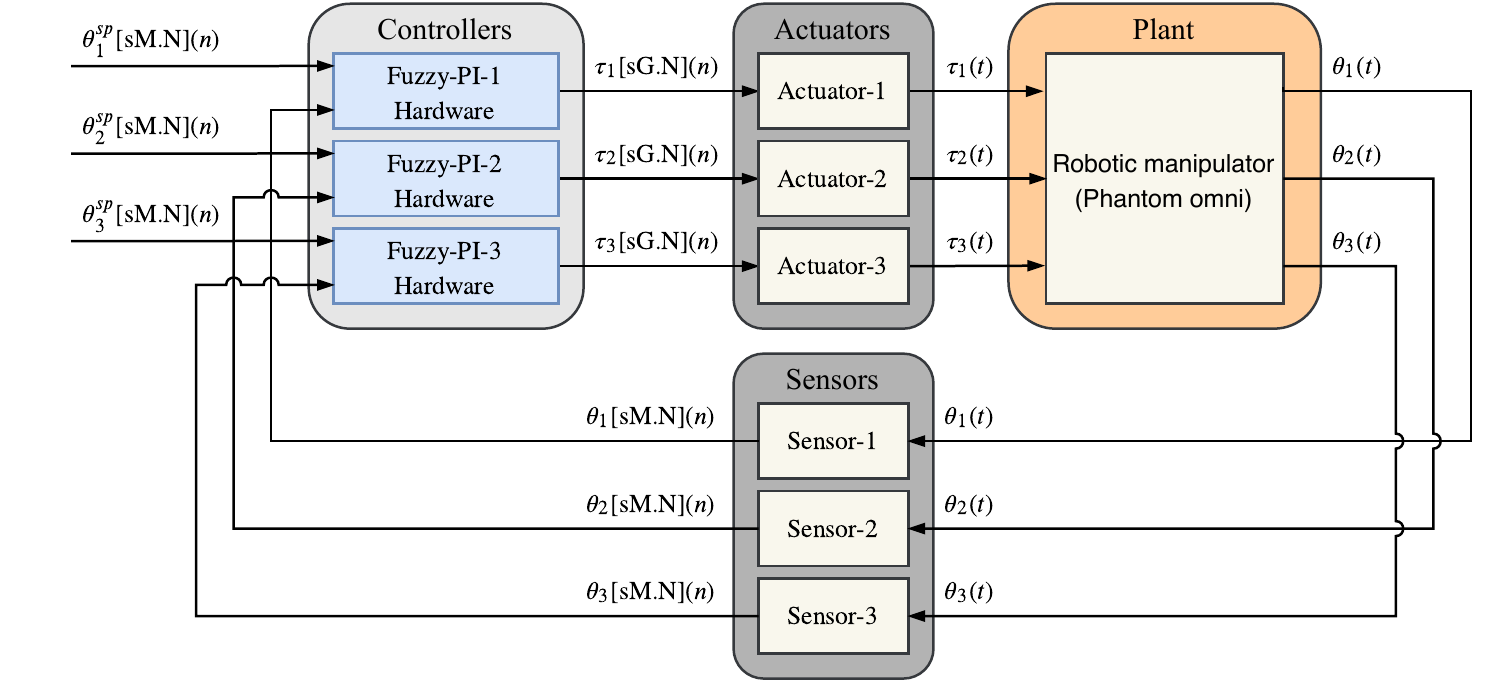}
		\caption{Simulated system used to validate the Fuzzy-PI hardware proposal. The plant is the 3-DOF Phantom Omni robotic manipulator and there are three Fuzzy-PI hardware running in parallel.} 
		\label{FuzzyPIRobotControl}
	\end{center}
\end{figure}

Figure \ref{FuzzyPIRobotControl} shows the simulated system where the plant is the 3-DOF Phantom Omni robotic manipulator. The controlled variables are the angular position of the joints $\theta_1$, $\theta_2$ and $\theta_3$ and the actuator variables are the torques $\tau_1$, $\tau_2$ and $\tau_3$. The control system has three angular position sensors and each $i$-th $\text{Sensor-}i$ convert the $i$-th continuous angle signal, $\theta_i(t)$ to discrete angle signal, $\theta_i(n)$. There are three Fuzzy-PI hardware running in parallel and every $i$-th $\text{Sensor-}i$ is connected with a Fuzzy-PI hardware, $\text{Fuzzy-PI-}i$. Each $i$-th $\text{Fuzzy-PI-}i$ hardware generates the $i$-th discrete torques acting signal, $\tau_i(n)$, and every $i$-th discrete torque signal, $\tau_i(n)$, is connected to $i$-th actuator, $\text{Actuator-}i$. Finally, each $i$-th actuator, $\text{Actuator-}i$, generates the $i$-th continuous torque signal, $\tau_i(t)$ to the applied on the robotic manipulator. The set point variables (or reference signal) are angular position of the joints and they are expressed by $\theta^{sp}_1(n)$, $\theta^{sp}_2(n)$ and $\theta^{sp}_3(n)$.

\begin{figure}[ht]
	\begin{center}
		\includegraphics[width=1\textwidth]{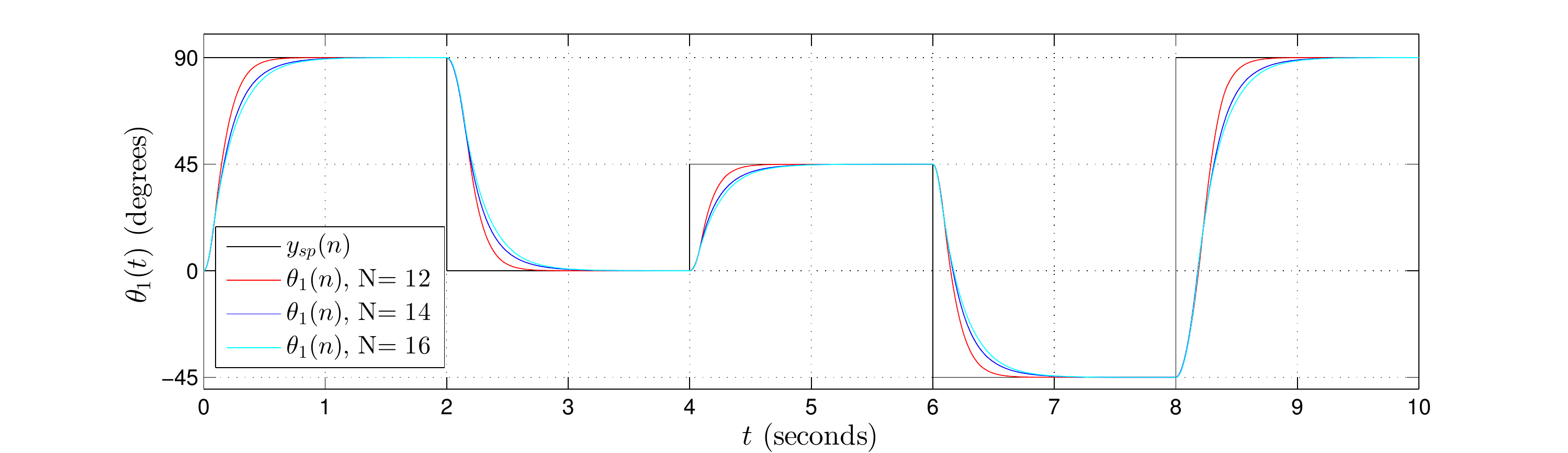}
		\caption{Validation results from the proposed Takagi-Sugeno Fuzzy-PI hardware. Simulation trajectory for $\theta_1(t)$ with $\theta_1(n)$ using $N = \{12, 14, 16\}$ bits in the fractional part.} 
		\label{FigResultsTheta1}
	\end{center}
\end{figure}

\begin{figure}[ht]
	\begin{center}
		\includegraphics[width=1\textwidth]{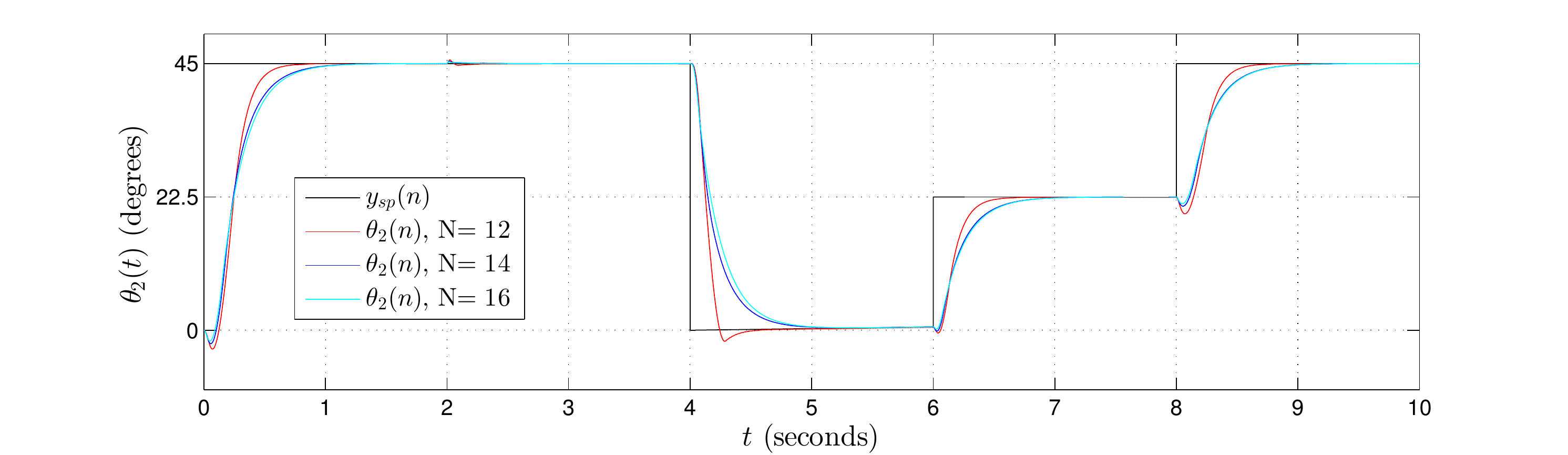}
		\caption{Validation results from the proposed Takagi-Sugeno Fuzzy-PI hardware. Simulation trajectory for $\theta_2(t)$ with $\theta_2(n)$ using $N = \{12, 14, 16\}$ bits in the fractional part.} 
		\label{FigResultsTheta2}
	\end{center}
\end{figure}

\begin{figure}[ht]
	\begin{center}
		\includegraphics[width=1\textwidth]{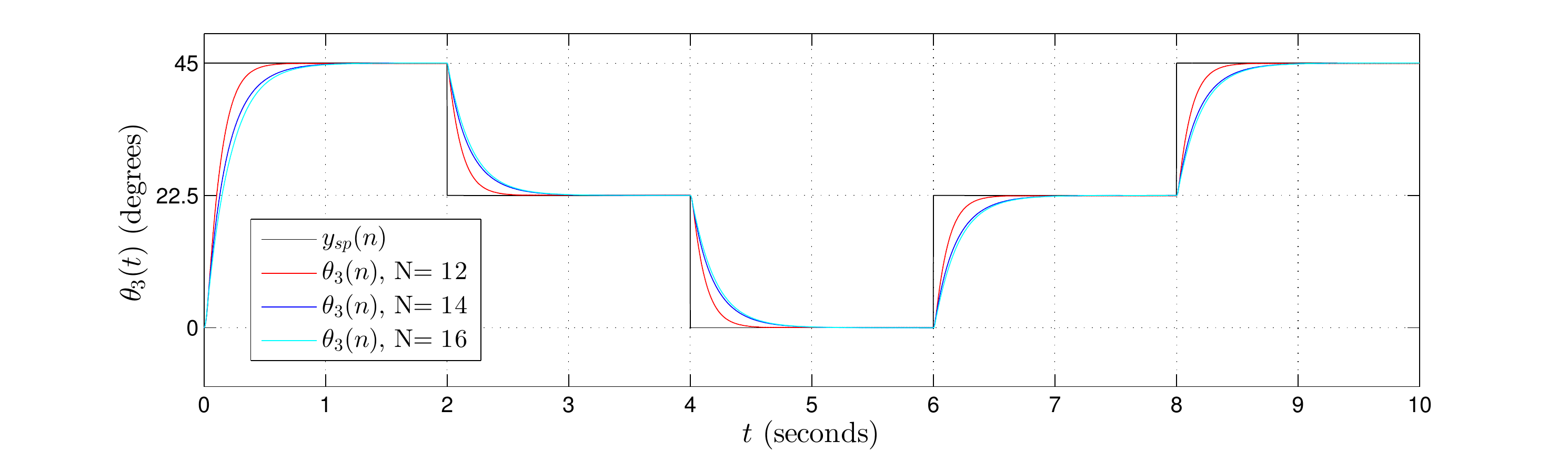}
		\caption{Validation results from the proposed Takagi-Sugeno Fuzzy-PI hardware. Simulation trajectory for $\theta_3(t)$ with $\theta_3(n)$ using $N = \{12, 14, 16\}$ bits in the fractional part.} 
		\label{FigResultsTheta3}
	\end{center}
\end{figure}

Figures \ref{FigResultsTheta1}, \ref{FigResultsTheta2} and \ref{FigResultsTheta3} present the hardware validation results for various resolutions in terms of the number of bits of the fractional part, $ N = \{12, 14, 16\}$ for discrete controlled variables $\theta_1(n)$, $\theta_2(n)$ and $\theta_3(n)$, respectively. The simulation trajectory was of $10$ seconds and every $2$ seconds was changing. Table \ref{TabTrajectoryChanging} shows the angle trajectory changing for set point variables $\theta^{sp}_1(n)$, $\theta^{sp}_2(n)$ and $\theta^{sp}_3(n)$. Simulations used $t_s=1 \times 10^{-5}$, $\text{Kp}=2000$ and $\text{Ki}=0.1$ for each $i$-th $\text{Fuzzy-PI-}i$ hardware.

\begin{table}[ht]
	\small
	\begin{center}
		\caption{Angle trajectory changing for set point variables $\theta^{sp}_1(n)$, $\theta^{sp}_2(n)$ and $\theta^{sp}_3(n)$.}
		\begin{tabular}{cccccc}
			\hline
			Set point & $0-2 \, \text{s}$ &  $2 \, \text{s}-4 \, \text{s}$ & $4 \, \text{s}-6 \, \text{s}$ & $6 \, \text{s}-8 \, \text{s}$ & $8 \, \text{s}-10 \, \text{s}$ \\
			\hline \noalign{\smallskip} 
	           $\theta^{sp}_1(n)$ (Figure \ref{FigResultsTheta1}) & $90\degree$ & $0\degree$ & $45\degree$ & $-45\degree$ & $90\degree$  \\[.1cm]
	           $\theta^{sp}_2(n)$ (Figure \ref{FigResultsTheta2})& $45\degree$ & $45\degree$ & $0\degree$ & $22.5\degree$ & $45\degree$  \\[.1cm]
	           $\theta^{sp}_3(n)$ (Figure \ref{FigResultsTheta3})& $45\degree$ & $22.5\degree$ & $0\degree$ & $22.5\degree$ & $45\degree$  \\
			\hline        
		\end{tabular}
		\label{TabTrajectoryChanging}
	\end{center}
\end{table}

In the results presented in Figures \ref{FigResultsTheta1}, \ref{FigResultsTheta2} and \ref{FigResultsTheta3} it is possible to observed that the controller followed the plant reference in all cases. Results also showed that the Takagi-Sugeno Fuzzy-PI hardware proposal has been following the reference even for a small amount of bits, that is, a low resolution.

\section{Comparison with other works}

\subsection{Throughput comparison}

Table \ref{TabComparisonThroughput} shows a comparison with other works in the literature. Parameters like inference machine (IM) type (Takagi-Sugeno or Mamdani), number of inputs (NI), number of rules (NR), number of outputs (NO), number of bits (NB), throughput in Msps, $R_s$ and Mflips (Mega fuzzy logic inference per second) are showed. In additional, Table \ref{TabComparisonThroughput} also shows the speedups (in Msps and Mflips) achieved of the TS-FIMM-OS, TS-FIMM-P, Fuzzy-PI controller with TS-FIMM-OS (Fuzzy-PI-OS) and with TS-FIMM-P (Fuzzy-PI-P) over the other works in the literature. The value in flips can be calculated as $\text{NR} \times R_s$.

In the work presented in \cite{PaperFPGAFuzzySugenoControl7A}, the results were obtained for several cases and for one with two inputs, $35$ rules and one output (vehicle parking problem) the proposed hardware achieved a maximum clock about $66.251 \, \text{MHz}$ with $10$ bits \cite{PaperFPGAFuzzySugenoControl7B, PaperFPGAFuzzySugenoControl7C}. However, the FIM takes $10$ clocks to complete the inference step; in other words, the hardware proposal in \cite{PaperFPGAFuzzySugenoControl7A} achieves a throughput in Msps of about $\frac{66.251}{10} \approx 6.63 \, \text{Msps}$ and in Mflips of about  $6.63 \times 35 \approx 232.05 \, \text{Mflips}$. The speedup in Msps for the TS-FIMM-OS, TS-FIMM-P, Fuzzy-PI-OS and Fuzzy-PI-P are $\frac{12.05 \, \text{Msps}}{6.63 \, \text{Msps}} \approx 1.82$, $\frac{17.63 \, \text{Msps}}{6.63\, \text{Msps}}\approx 2.66$,  $\frac{10.16 \, \text{Msps}}{6.63\, \text{Msps}}\approx1.53$, and $\frac{13.86 \, \text{Msps}}{6.63\, \text{Msps}}\approx 2.09$, respectively. As the hardware proposal in this paper used $49$ rules, the speedup in Mflips can be calculated as the throughput in Msps $\times \frac{49}{35}$, that is, the speedup for the TS-FIMM-OS, TS-FIMM-P, Fuzzy-PI-OS and Fuzzy-PI-P are $1.82 \times 1.4 \approx 2.55$, $1.82 \times 1.4 \approx 3.72$, $1.53 \times 1.4 \approx 2.14$, and $2.09 \times 1.4 \approx 2.93$, respectively.

The work presented in \cite{PaperFPGAFuzzySugenoControl1} proposes a Takagi-Sugeno fuzzy controller on FPGA with  two inputs, $6$ rules and three outputs. The hardware achieved a throughput of about $1 \, \text{Msps}$ with $8$ bits on the bus. With $8$ bits, the speedup in Msps for the TS-FIMM-OS, TS-FIMM-P, Fuzzy-PI-OS and Fuzzy-PI-P are $\frac{11.94 \, \text{Msps}}{1 \, \text{Msps}} \approx 11.94$, $\frac{17.55\, \text{Msps}}{1\, \text{Msps}}\approx 17.55$,  $\frac{10.77 \, \text{Msps}}{1\, \text{Msps}}\approx10.77$, and $\frac{15.13 \, \text{Msps}}{1\, \text{Msps}}\approx 15.13$, respectively. The speedup in Mflips is about $\frac{49}{6} \approx  8.16\times$ over the speedup in Msps.  

In \cite{PaperFuzzyControlFPGA3}, a Mamdani fuzzy logic controller on FPGA was proposed. The hardware carries out a throughput of about $25 \, \text{Mflips}$ with two inputs, $49$ rules, one output, and $16$ bits. Using $16$ bits, the speedup in Mflips for the TS-FIMM-OS, TS-FIMM-P, Fuzzy-PI-OS and Fuzzy-PI-P are $\frac{11.28 \times 49 \, \text{Mflips}}{25 \, \text{Mflips}} \approx 22.11$, $\frac{16.98 \times 49\, \text{Mflips}}{25\, \text{Mflips}}\approx 33.28$,  $\frac{9.59 \times 49 \, \text{Mflips}}{25\, \text{Mflips}}\approx18.79$, and $\frac{13.41 \times 49 \, \text{Mflips}}{25\, \text{Mflips}}\approx 26.28$, respectively. As the number of rules is $49$, the speedup in Msps is equal to Mflips.

The work presented in \cite{PaperFPGAFuzzy4} uses a Mamdani inference machine and the throughput in Mflips is about $48.23\, \text{Mflips}$. The hardware designed in \cite{PaperFPGAFuzzy4} operated with $8$ bits, four inputs, $9$ rules and one output. The speedup in Mflips, with $8$ bits, for the TS-FIMM-OS, TS-FIMM-P, Fuzzy-PI-OS and Fuzzy-PI-P are $\frac{11.94 \times 49 \, \text{Mflips}}{48.23 \, \text{Mflips}} \approx 12.13$, $\frac{17.55 \times 49\, \text{Mflips}}{48.23\, \text{Mflips}}\approx 17.83$,  $\frac{10.77 \times 49 \, \text{Mflips}}{48.23\, \text{Mflips}}\approx10.94$, and $\frac{13.41 \times 49 \, \text{Mflips}}{48.23\, \text{Mflips}}\approx 15.37$, respectively. The speedup in Msps is about $\frac{9}{49} \approx  0.18\times$ over the speedup in Mflips.  

The hardware used in \cite{PaperFuzzyControlFPGA1} takes $6$ clocks cycles over $10 \, \text{MHz}$ (in four states) to execute a M-IM with $16$ bits. This is equivalent to a throughput of about $\frac{10 \, \text{MHz}}{6} \approx 1.67 \, \text{Msps}$. The scheme proposed in \cite{PaperFuzzyControlFPGA1} used two inputs, $25$ rules and one output. The speedup in Msps for the TS-FIMM-OS, TS-FIMM-P, Fuzzy-PI-OS and Fuzzy-PI-P are $\frac{11.28 \, \text{Msps}}{1.67 \, \text{Msps}} \approx 6.75$, $\frac{16.98 \, \text{Msps}}{1.67\, \text{Msps}}\approx 10.17$,  $\frac{9.59 \, \text{Msps}}{1.67\, \text{Msps}}\approx5.74$, and $\frac{13.41 \, \text{Msps}}{1.67 \, \text{Msps}}\approx 8.03$, respectively. The speedup in Mflips is about $\frac{49}{25} \approx 1.96\times$ over the speedup in Msps.  

The works presented in \cite{PaperFuzzyControlFPGA6, PaperFuzzyControlFPGA8} shows a hardware can achieve about $1\, \text{Msps}$. The work presented in \cite{PaperFuzzyControlFPGA6} uses two inputs, $25$ rules, one output and $8$ bits and the designer presented in \cite{PaperFuzzyControlFPGA8} was projected with three inputs, $42$ rules and one output. The speedup in Msps for the TS-FIMM-OS, TS-FIMM-P, Fuzzy-PI-OS and Fuzzy-PI-P are equal to previously calculated values used in \cite{PaperFPGAFuzzySugenoControl1}. The speedup in Mflips are about $\frac{49}{25} \approx 1.96\times$ and $\frac{49}{42} \approx 1.16\times$ over the speedup in Msps for works  \cite{PaperFuzzyControlFPGA6} and \cite{PaperFuzzyControlFPGA8}, respectively.  

Finally, the hardware proposes in \cite{PaperFPGAFuzzySugenoControl3} achieved a throughput of about $1.56\, \text{Msps}$ with three inputs, two outputs and $24$ bits. The speedup in Msps for the TS-FIMM-OS, TS-FIMM-P, Fuzzy-PI-OS and Fuzzy-PI-P are $\frac{11.28 \, \text{Msps}}{1.56 \, \text{Msps}} \approx 7.23$, $\frac{16.98 \, \text{Msps}}{1.56\, \text{Msps}}\approx 10.88$,  $\frac{9.59 \, \text{Msps}}{1.56\, \text{Msps}}\approx 6.15$, and $\frac{13.41 \, \text{Msps}}{1.56 \, \text{Msps}}\approx 8.59 $, respectively. The fuzzy system proposed in \cite{PaperFPGAFuzzySugenoControl3} does not use linguistic fuzzy rules and it cannot calculate the throughput in Mflips.

\begin{table}[ht]
	\small
	\begin{center}
		\caption{Throughput comparison with other works.}
		\begin{tabular}{ccccccccccc}
			\hline
			\multirow{2}{*}{References} & \multirow{2}{*}{IM} & \multirow{2}{*}{NI} & \multirow{2}{*}{NR} & \multirow{2}{*}{NO} & \multirow{2}{*}{NB} &  \multirow{2}{*}{Msps} & \multirow{2}{*}{Mflips} & \multirow{2}{*}{This work} &\multicolumn{2}{c}{Speedup}\\         
               & & &  &  &  &   &  && Msps & Mflips \\                  
			\hline
               \multirow{4}{*}{\cite{PaperFPGAFuzzySugenoControl7A} (2013)} & \multirow{4}{*}{TS-IM} & \multirow{4}{*}{$2$} & \multirow{4}{*}{$35$} & \multirow{4}{*}{$1$} & \multirow{4}{*}{$10$} & \multirow{4}{*}{$\approx 6.63$} &  \multirow{4}{*}{$\approx 232.05$}&TS-FIMM-OS & $\approx 1.82\times$ & $\approx 2.55\times$ \\
               & & &  &  &  &   & &TS-FIMM-P & $\approx 2.66\times$ & $\approx 3.72\times$  \\    
               & & &  &  &  &   & &Fuzzy-PI-OS & $\approx 1.53\times$ &  $\approx 2.14\times$\\    
              & & &  &  &  &   &  &Fuzzy-PI-P & $\approx 2.09\times$ &  $\approx2.93\times$\\    
              \hline
               \multirow{4}{*}{\cite{PaperFPGAFuzzySugenoControl1} (2014) } & \multirow{4}{*}{TS-IM} & \multirow{4}{*}{$2$} & \multirow{4}{*}{$6$} & \multirow{4}{*}{$3$} & \multirow{4}{*}{$8$} & \multirow{4}{*}{$\approx 1.00$} &  \multirow{4}{*}{$\approx 6.00$}&TS-FIMM-OS & $\approx 11.94\times$ & $\approx 97.43\times$ \\
               & & &  &  &  &   & &TS-FIMM-P & $\approx 17.55\times$ & $\approx 143.20\times$  \\    
               & & &  &  &  &   & &Fuzzy-PI-OS & $\approx 10.77\times$ &  $\approx 87.88\times$\\    
              & & &  &  &  &   &  &Fuzzy-PI-P & $\approx 15.13\times$ &  $\approx123.46\times$\\    
              \hline
               \multirow{4}{*}{\cite{PaperFuzzyControlFPGA3} (2015)} & \multirow{4}{*}{M-IM} & \multirow{4}{*}{$2$} & \multirow{4}{*}{$49$} & \multirow{4}{*}{$1$} & \multirow{4}{*}{$16$} & \multirow{4}{*}{$\approx 0.51$} &  \multirow{4}{*}{$\approx 25.00$}&TS-FIMM-OS & $\approx 22.11\times$ & $\approx 22.11\times$ \\
               & & &  &  &  &   & &TS-FIMM-P & $\approx  33.28\times$ & $\approx  33.28\times$  \\    
               & & &  &  &  &   & &Fuzzy-PI-OS & $\approx 18.79\times$ &  $\approx 18.79\times$\\    
              & & &  &  &  &   &  &Fuzzy-PI-P & $\approx 26.28\times$ &  $\approx26.28\times$\\    
              \hline
               \multirow{4}{*}{\cite{PaperFPGAFuzzy4} (2016)} & \multirow{4}{*}{M-IM} & \multirow{4}{*}{$4$} & \multirow{4}{*}{$9$} & \multirow{4}{*}{$1$} & \multirow{4}{*}{$8$} & \multirow{4}{*}{$\approx 5.36$} &  \multirow{4}{*}{$\approx 48.23$}&TS-FIMM-OS & $\approx 2.18\times$ & $\approx 12.13\times$ \\
               & & &  &  &  &   & &TS-FIMM-P & $\approx  3.20\times$ & $\approx  17.83\times$  \\    
               & & &  &  &  &   & &Fuzzy-PI-OS & $\approx 1.97\times$ &  $\approx 10.94\times$\\    
              & & &  &  &  &   &  &Fuzzy-PI-P & $\approx 2.76\times$ &  $\approx15.37\times$\\    
              \hline
               \multirow{4}{*}{\cite{PaperFuzzyControlFPGA1} (2018)} & \multirow{4}{*}{M-IM} & \multirow{4}{*}{$2$} & \multirow{4}{*}{$25$} & \multirow{4}{*}{$1$} & \multirow{4}{*}{$16$} & \multirow{4}{*}{$\approx 1.67$} &  \multirow{4}{*}{$\approx 41.75$}&TS-FIMM-OS & $\approx 6.75 \times$ & $\approx 13.23\times$ \\
               & & &  &  &  &   & &TS-FIMM-P & $\approx  10.17 \times$ & $\approx  19.93\times$  \\    
               & & &  &  &  &   & &Fuzzy-PI-OS & $\approx 5.74\times$ &  $\approx 11.25\times$\\    
              & & &  &  &  &   &  &Fuzzy-PI-P & $\approx 8.03\times$ &  $\approx 15.74\times$\\    
              \hline
               \multirow{4}{*}{\cite{PaperFuzzyControlFPGA6} (2019)} & \multirow{4}{*}{M-IM} & \multirow{4}{*}{$2$} & \multirow{4}{*}{$25$} & \multirow{4}{*}{$1$} & \multirow{4}{*}{$8$} & \multirow{4}{*}{$\approx 1.00$} &  \multirow{4}{*}{$\approx 25.00$}&TS-FIMM-OS & $\approx 11.94\times$ & $\approx 23.40\times$ \\
               & & &  &  &  &   & &TS-FIMM-P & $\approx 17.55\times$ & $\approx  34.40\times$  \\    
               & & &  &  &  &   & &Fuzzy-PI-OS & $\approx 10.77\times$ &  $\approx 21.11\times$\\    
              & & &  &  &  &  & & Fuzzy-PI-P & $\approx 15.13 \times$ &  $\approx 29.65\times$\\    
              \hline
               \multirow{4}{*}{\cite{PaperFuzzyControlFPGA8} (2019)} & \multirow{4}{*}{M-IM} & \multirow{4}{*}{$3$} & \multirow{4}{*}{$42$} & \multirow{4}{*}{$1$} & \multirow{4}{*}{$-$} & \multirow{4}{*}{$\approx 1.00$} &  \multirow{4}{*}{$\approx 42.00$}&TS-FIMM-OS & $\approx 11.94\times$ & $\approx 13.85\times$ \\
               & & &  &  &  &   & &TS-FIMM-P & $\approx 17.55\times$ & $\approx  20.36\times$  \\    
               & & &  &  &  &   & &Fuzzy-PI-OS & $\approx 10.77\times$ &  $\approx 12.49\times$\\    
              & & &  &  &  &  & & Fuzzy-PI-P & $\approx 15.13 \times$ &  $\approx 17.55\times$\\   
              \hline
               \multirow{4}{*}{\cite{PaperFPGAFuzzySugenoControl3} (2019)} & \multirow{4}{*}{TS-IM} & \multirow{4}{*}{$3$} & \multirow{4}{*}{$-$} & \multirow{4}{*}{$2$} & \multirow{4}{*}{$24$} & \multirow{4}{*}{$\approx 1.56$} &  \multirow{4}{*}{$-$}&TS-FIMM-OS & $\approx 7.23 \times$ & $-$ \\
               & & &  &  &  &   & &TS-FIMM-P & $\approx  10.88 \times$ & $-$  \\    
               & & &  &  &  &   & &Fuzzy-PI-OS & $\approx 6.15 \times$ &  $-$\\    
              & & &  &  &  &   & & Fuzzy-PI-P & $\approx 8.59 \times$ &  $-$\\    
              \hline
		\end{tabular}
		\label{TabComparisonThroughput}
	\end{center}
\end{table}

\subsection{Hardware occupation comparison }

Table \ref{TabComparisonOccupation} shows a comparison regarding the hardware occupation between the proposed hardware in this work and other literature works presented in Table \ref{TabComparisonThroughput}. The second, third, fourth and fifth columns show the type of FPGA, the number of logic cells (NLC), the number of multipliers (NMULT) and the number of bits in memory block RAMs (NBitsM), respectively and the last three columns show the ratio of the hardware occupation between the proposal presented here, $N^{\text{work}}_{\text{hardware}}$, and literature works, $N^{\text{ref}}_{\text{hardware}}$, presented in Table \ref{TabComparisonThroughput}. The ratio of the hardware occupation can be expressed as  
\begin{equation}
R_{\text{occupation}} = 
\left\{\begin{matrix}
\frac{N^{\text{work}}_{\text{hardware}}}{N^{\text{ref}}_{\text{hardware}}} & \text{, for }  N^{\text{work}}_{\text{hardware}}>0 \text{ and } N^{\text{ref}}_{\text{hardware}}>0\\ 
& \\
\frac{1}{N^{\text{ref}}_{\text{hardware}}} & \text{, for }  N^{\text{work}}_{\text{hardware}}=0 \text{ and } N^{\text{ref}}_{\text{hardware}}>0 \\
& \\ 
N^{\text{work}}_{\text{hardware}} &  \text{, for }  N^{\text{work}}_{\text{hardware}}>0 \text{ and } N^{\text{ref}}_{\text{hardware}}=0\\ 
& \\
1 & \text{, for }  N^{\text{work}}_{\text{hardware}}=0 \text{ and } N^{\text{ref}}_{\text{hardware}}=0 
\end{matrix}\right. ,
\end{equation}
where $N^{\text{work}}_{\text{hardware}}$ and $N^{\text{ref}}_{\text{hardware}}$ can be replaced by NLC, NMULT or NBitsM.

The work presented in \cite{PaperFPGAFuzzySugenoControl7A} used a Spartan 3A DSP FPGA from Xilinx and it has a hardware occupation of about $199$ slices, $4$ multipliers and $1$ block RAM. As this FPGA uses about $2.25$ LC per slice, it used about $447$ LC and it has $1512 \, \text{K}$ bits per block RAM. The scheme proposed in \cite{PaperFPGAFuzzySugenoControl1} used a Cyclone II EP2C35F672C6 FPGA from Intel and it has a hardware occupation of about $1622$ logic cells and $8.19$ Kbits of memory. The EP2C35 FPGA has 105 block RAM and $4\text{,}096$ memory bits per block ($4\text{,}608$ bits per block including $512$ parity bits).

In \cite{PaperFuzzyControlFPGA3}, the work assign a Arria V GX 5AGXFB3H4F40C5NES FPGA from Intel and it has a hardware occupation of about $3248$ ALMs and $6.592$ Kbits of memory. The Arria V GX 5AGX has two combinational logic cells per ALM. The hardware proposed in \cite{PaperFPGAFuzzy4} employs a Spartan 6 FPGA from Xilinx and it has a hardware occupation of about $544$ LUTs and $32$ multipliers. As this FPGA uses about $1.6$ LC per LUT, it used about $447$ LC.

The hardware presented in the manuscript \cite{PaperFuzzyControlFPGA1} utilizes a Spartan 6 FPGA from Xilinx and it has a hardware occupation of about $1802$ slices and $5$ multipliers. As this FPGA works with $6.34$ LC per slice, it used about $11425$ LC. The proposal described in \cite{PaperFuzzyControlFPGA8} take advantage of Virtex 5 xc5vfx70t-3ff1136 FPGA from Xilinx and it has a hardware occupation of about $8195$ LUTs and $53$ multipliers. As this FPGA uses about $1.6$ LC per LUT, it used about $13108$ LC. 6-input LUT, they use the multiplier 1.6. The work presented in \cite{PaperFPGAFuzzySugenoControl3} used a Virtex 7 VX485T-2 FPGA from Xilinx and it has a hardware occupation of about $1948$ slices and $38$ multipliers. As this FPGA uses about $6.4$ LC per slice, it used about $12468$ LC.

\begin{table}[ht]
	\small
	\begin{center}
		\caption{Hardware occupation comparison with other works.}
		\begin{tabular}{ccccccccc}
			\hline
			\multirow{2}{*}{References} & \multirow{2}{*}{FPGA} & \multirow{2}{*}{NLC} & \multirow{2}{*}{NMULT} & \multirow{2}{*}{NBitsM} & \multirow{2}{*}{This work} & \multicolumn{3}{c}{$R_{\text{occupation}}$}\\         
               & & &  &  &  &    NLC & NMULT & NBitsM \\                  
			\hline
               \multirow{4}{*}{\cite{PaperFPGAFuzzySugenoControl7A} (2013)} & \multirow{4}{*}{Spartan 3A} & \multirow{4}{*}{$447$} & \multirow{4}{*}{$4$} & \multirow{4}{*}{$1512 \, \text{K}$} & TS-FIMM-OS & $\approx 26.24 \times$ & \multirow{4}{*}{$\approx 12.25 \times$} &\multirow{4}{*}{$\approx 10^{-6}\times$} \\
 & & &  &  & TS-FIMM-P& $\approx 22.61 \times$ &  & \\
 & & &  &  & Fuzzy-PI-OS & $\approx 26.24 \times$ &  &\\
 & & &  &  & Fuzzy-PI-P & $\approx 22.61 \times$ &  &  \\
              \hline
               \multirow{4}{*}{\cite{PaperFPGAFuzzySugenoControl1} (2014) } & \multirow{4}{*}{Cyclone II} & \multirow{4}{*}{$1622$} & \multirow{4}{*}{$0$} & \multirow{4}{*}{$8.19 \, \text{K}$} & TS-FIMM-OS & $\approx 6.51 \times$ & \multirow{4}{*}{$49 \times$} &\multirow{4}{*}{$\approx 10^{-3}\times$} \\
 &  & &  &  & TS-FIMM-P& $\approx 5.51 \times$ &  &  \\
 & & &  &  & Fuzzy-PI-OS & $\approx 6.74 \times$ & & \\
 & & &  &  & Fuzzy-PI-P & $\approx 5.75 \times$ &  &  \\
              \hline
               \multirow{4}{*}{\cite{PaperFuzzyControlFPGA3} (2015)} & \multirow{4}{*}{Arria V GX} & \multirow{4}{*}{$6496$} & \multirow{4}{*}{$0$} & \multirow{4}{*}{$6.592 \, \text{K}$} & TS-FIMM-OS & $\approx 2.53 \times$ &  \multirow{4}{*}{$49 \times$} &\multirow{4}{*}{$\approx 10^{-3}\times$} \\
 &  & &  &  & TS-FIMM-P& $\approx 2.21 \times$ &  &  \\
 & & &  &  & Fuzzy-PI-OS & $\approx 2.61 \times$ &  &  \\
 & & &  &  & Fuzzy-PI-P & $\approx 2.29 \times$ &  &  \\
              \hline
               \multirow{4}{*}{\cite{PaperFPGAFuzzy4} (2016)} & \multirow{4}{*}{Spartan 6} & \multirow{4}{*}{$871$} & \multirow{4}{*}{$32$} & \multirow{4}{*}{$0 \, \text{K}$} & TS-FIMM-OS & $\approx 12.13 \times$ & \multirow{4}{*}{$\approx 1.53 \times$} & \multirow{4}{*}{$ 1 \times$}\\
 & & &  &  & TS-FIMM-P  & $\approx 10.28 \times$ &  &  \\
 & & &  &  & Fuzzy-PI-OS & $\approx 12.56 \times$ &  &  \\
 & & &  &  & Fuzzy-PI-P   & $\approx 10.71\times$ &  &  \\
              \hline
               \multirow{4}{*}{\cite{PaperFuzzyControlFPGA1} (2018)} & \multirow{4}{*}{Spartan 6} & \multirow{4}{*}{$11425$} & \multirow{4}{*}{$5$} & \multirow{4}{*}{$0 \, \text{K}$} & TS-FIMM-OS & $\approx 1.44 \times$ & \multirow{4}{*}{$\approx 9.8 \times$} & \multirow{4}{*}{$ 1 \times$}\\
 & & &  &  & TS-FIMM-P  & $\approx 1.25 \times$ &  &  \\
 & & &  &  & Fuzzy-PI-OS & $\approx 1.48 \times$ &  &  \\
 & & &  &  & Fuzzy-PI-P   & $\approx 1.30\times$ &  &  \\
              \hline
               \multirow{4}{*}{\cite{PaperFuzzyControlFPGA8} (2019)} & \multirow{4}{*}{Virtex 5} & \multirow{4}{*}{$13108$} & \multirow{4}{*}{$53$} & \multirow{4}{*}{$0 \, \text{K}$} & TS-FIMM-OS & $\approx 1.25 \times$ & \multirow{4}{*}{$\approx 0.93 \times$} & \multirow{4}{*}{$ 1 \times$}\\
 & & &  &  & TS-FIMM-P  & $\approx 1.09 \times$ &  &  \\
 & & &  &  & Fuzzy-PI-OS & $\approx 1.29 \times$ &  &  \\
 & & &  &  & Fuzzy-PI-P   & $\approx 1.13\times$ &  &  \\
              \hline
               \multirow{4}{*}{\cite{PaperFPGAFuzzySugenoControl3} (2019)} & \multirow{4}{*}{Virtex 7} & \multirow{4}{*}{$12468$} & \multirow{4}{*}{$38$} & \multirow{4}{*}{$0 \, \text{K}$} & TS-FIMM-OS & $\approx 1.32 \times$ & \multirow{4}{*}{$\approx 1.29 \times$} & \multirow{4}{*}{$ 1 \times$}\\
 & & &  &  & TS-FIMM-P  & $\approx 1.15 \times$ &  &  \\
 & & &  &  & Fuzzy-PI-OS & $\approx 1.36 \times$ &  &  \\
 & & &  &  & Fuzzy-PI-P   & $\approx 1.19\times$ &  &  \\
              \hline
		\end{tabular}
		\label{TabComparisonOccupation}
	\end{center}
\end{table}

\subsection{Power consumption comparison}

Table \ref{TabComparisonDynamicPower} shows the dynamic power saving regards the dynamic power. The dynamic power can be expressed as
\begin{equation}\label{DynamicPower1}
P_d \propto N_g \times F_{\text{clk} }\times V_{DD}^2,
\end{equation}
where $N_g$ is the number of elements (or gates), $F_{\text{clk}}$ is the maximum clock frequency and $V_{DD}$ is the supply voltage. The frequency dependence is more severe than equation \ref{DynamicPower1} suggests, given that the frequency at which a CMOS circuit can operate is approximately proportional to the voltage \cite{MCCOOL201239}. Thus, the dynamic power can be expressed as
\begin{equation}\label{DynamicPower2}
P_d \propto N_g \times F_{\text{clk}}^3.
\end{equation}
For all comparisons, the number of elements, $N_g$, was calculated as
\begin{equation}\label{DynamicPower1}
N_g = \text{NLC} + \text{NMULT}.
\end{equation}

Based on Equation \ref{DynamicPower2}, the dynamic power saving can be expressed as
\begin{equation}\label{DynamicPower3}
S_d = \frac{N^{\text{ref}}_g \times \left(F^{\text{ref}}_{\text{clk}}\right)^3}{N^{\text{work}}_g \times \left(F^{\text{work}}_{\text{clk}}\right)^3},
\end{equation}
where the $N^{\text{ref}}_g$ and $F^{\text{ref}}_{\text{clk}}$ are the number of elements ($\text{NLC} + \text{NMULT}$) and the maximum clock frequency of the literature works, respectively and the $N^{\text{work}}_g$ and $F^{\text{work}}_{\text{clk}}$ are the number of elements ($\text{NLC} + \text{NMULT}$) and the maximum clock frequency of this work, respectively. Differently from the literature, the hardware proposed here uses a fully parallelization layout, and it spends a one clock cycle per sample processing. In other words, the maximum clock frequency is equivalent to the throughput, $F^{\text{work}}_{\text{clk}} \equiv R_s$.

\begin{table}[ht]
	\small
	\begin{center}
		\caption{Dynamic power comparison with other works.}
		\begin{tabular}{cccccccc}
			\hline
			\multirow{2}{*}{References} & \multirow{2}{*}{FPGA} & \multirow{2}{*}{$N^{\text{ref}}_g$} & \multirow{2}{*}{$F^{\text{ref}}_{\text{clk}}$ (MHz)} & \multirow{2}{*}{This work} &  \multirow{2}{*}{$N^{\text{work}}_g$} & \multirow{2}{*}{$F^{\text{work}}_{\text{clk}}$ (MHz)} & \multirow{2}{*}{$S_d$} \\          
        & & &  & &  &  & \\           
			\hline
               \multirow{4}{*}{\cite{PaperFPGAFuzzySugenoControl7A} (2013)} & \multirow{4}{*}{Spartan 3A} & \multirow{4}{*}{$451$} & \multirow{4}{*}{$66.251$} & TS-FIMM-OS & $11779$  & \multirow{4}{*}{$6.63$} & $\approx 38.20 \times$ \\
 & & &  & TS-FIMM-P& $10157 $&  & $\approx 44.30 \times$\\
 & & &  & Fuzzy-PI-OS & $11779$ &  &$\approx 38.20 \times$\\
 & & &  &  Fuzzy-PI-P & $10157$  &  & $\approx 44.30 \times$ \\
              \hline
               \multirow{4}{*}{\cite{PaperFuzzyControlFPGA3} (2015)} & \multirow{4}{*}{Arria V GX} & \multirow{4}{*}{$6496$} & \multirow{4}{*}{$125$} & TS-FIMM-OS & $16453$  & \multirow{4}{*}{$0.51$} & \multirow{4}{*}{$\approx 10^6 \times$} \\
 & & &  & TS-FIMM-P& $14377$&  & \\
 & & &  & Fuzzy-PI-OS & $17001$ &  &\\
 & & &  &  Fuzzy-PI-P & $14926$  &  & \\
              \hline              
              \multirow{4}{*}{\cite{PaperFPGAFuzzy4} (2016)} & \multirow{4}{*}{Spartan 6} & \multirow{4}{*}{$903$} & \multirow{4}{*}{$20$} & TS-FIMM-OS & $6598$  & \multirow{4}{*}{$5.36$} & $\approx 4.42 \times$ \\
 & & &  & TS-FIMM-P& $5590$&  & $\approx 5.22 \times$\\
 & & &  & Fuzzy-PI-OS & $6834$ &  &$\approx 4.27 \times$\\
 & & &  &  Fuzzy-PI-P & $5826$  &  & $\approx 5.01 \times$ \\
              \hline
              \multirow{4}{*}{\cite{PaperFuzzyControlFPGA1} (2018)} & \multirow{4}{*}{Spartan 6} & \multirow{4}{*}{$11430$} & \multirow{4}{*}{$10$} & TS-FIMM-OS & $10252$  & \multirow{4}{*}{$1.67$} & $\approx 149.16 \times$ \\
 & & &  & TS-FIMM-P& $8955$&  & $\approx  170.70 \times$\\
 & & &  & Fuzzy-PI-OS & $10595$ &  &$\approx 144.35 \times$\\
 & & &  &  Fuzzy-PI-P & $9298$  &  & $\approx  164.42 \times$ \\
              \hline             
              \multirow{4}{*}{\cite{PaperFPGAFuzzySugenoControl3} (2019)} & \multirow{4}{*}{Virtex 7} & \multirow{4}{*}{$12506$} & \multirow{4}{*}{$150$} & TS-FIMM-OS & $10252$  & \multirow{4}{*}{$1.56$} & \multirow{4}{*}{$\approx 10^5 \times$} \\
 & & &  & TS-FIMM-P& $8955$&  & \\
 & & &  & Fuzzy-PI-OS & $10595$ &  &\\
 & & &  &  Fuzzy-PI-P & $9298$  &  & \\
              \hline                                     
		\end{tabular}
		\label{TabComparisonDynamicPower}
	\end{center}
\end{table}

\subsection{Analysis of the comparison}

Results presented in Tables \ref{TabComparisonThroughput} and \ref{TabComparisonDynamicPower} demonstrate that the fully parallelization strategy adopted here can achieve significant speedups and power consumption reductions. On the other hand, the fully parallelization scheme can increase the hardware consumption, see Table \ref{TabComparisonOccupation}. 

The mean value of speedup was about $10.89 \times$ in Msps and $30.89 \times$ in Mflips (see Table \ref{TabComparisonThroughput}) and this results are very expressive to big data and MMD applications \cite{fuzzyBig1, fuzzyBig2, fuzzyBig3}. High-throughput fuzzy controllers are also important to speed control systems such as tactile internet applications \cite{tactile, Tactile1RESRC}. 

This manuscript proposal has LC resource higher utilization than the literature proposals (Table \ref{TabComparisonOccupation}). The mean value regarding NLC utilization was about $6.89 \times$; in other words, the fuzzy hardware scheme proposed here has used $6.89 \times$ more LC than the literature proposals. In the case of multipliers (NMULT), the mean value of the additional hardware was about $17.69 \times$. Despite being large relative values, Tables \ref{Tab1Results}, \ref{Tab2Results}, \ref{Tab3Results} and \ref{Tab4Results} show that the fuzzy hardware proposals in this work expend no more than $7 \%$ of the FPGA resource. Another important aspect is the block RAM resource utilization (NBitsM). The fully parallel computing scheme proposed here, do not spend clock time to access information in block RAM and this can increase the throughput and decrease the power consumption (see references \cite{PaperFPGAFuzzySugenoControl7A}, \cite{PaperFPGAFuzzySugenoControl1} and \cite{PaperFuzzyControlFPGA3} in Tables \ref{TabComparisonThroughput}, \ref{TabComparisonOccupation} and \ref{TabComparisonDynamicPower}).

The fully parallel designer allows to execute many operations per clock period, and this reduces the clock frequency operation and increases the throughput. Due to the non-linear relationship with clock frequency operation (see Equation \ref{DynamicPower2}), this strategy permits a considerable reduction of the dynamic power consumption (see Table \ref{TabComparisonDynamicPower}). The results presented in Table \ref{TabComparisonDynamicPower} show that the power saving can achieve values from $4$ until $10^6$ times and these results are quite significant and enable the use of the proposed hardware here in several IoT applications.

\section{Conclusions}
This work aimed to develop a dedicated hardware for a fuzzy inference machine of the Takagi-Sugeno applied a Fuzzy-PI controller. The developed hardware used a fully parallel implementation with fixed-point and floating-point representation in distinct parts of the proposed scheme. All details of the implementation were presented as well as results for synthesis and bit-precision simulations. The synthesis results were performed for several bit size resolutions and showed that the proposed hardware is viable and can be used in applications with critical processing time requirements. Through the synthesis data, curves were generated to predict hardware consumption and throughput to untested bit values, in order to characterize the proposed hardware. In addition, comparison results concerning throughput, hardware occupation, and power saving with other literature proposals were presented.

\section*{Acknowledgments} The authors wish to acknowledge the financial support of the Coordena\c{c}\~ao de Aperfei\c{c}oamento de Pessoal de N\'ivel Superior (CAPES) for their financial support.

\section*{Conflicts of interest} The authors declare no conflict of interest.

\bibliography{PaperMain}

\end{document}